\documentclass[twocolumn,superscriptaddress,floatfix,preprintnumbers,letterpaper]{revtex4-1}
\usepackage{amsmath}
\usepackage{gensymb}
\usepackage{float}
\usepackage{array}
\usepackage{bm}% bold math
\usepackage[colorlinks, pdfborder={0 0 0}, plainpages=false]{hyperref}
\usepackage{graphicx}
\usepackage{xspace}
\usepackage[usenames,dvipsnames]{color}
\usepackage{amssymb}
\usepackage[normalem]{ulem} %for \sout
\usepackage{lineno}
\usepackage{tabularx}
\usepackage{letltxmacro}
\usepackage{amsfonts}
\usepackage{amsmath}
\usepackage{amssymb}
\usepackage{bm}
\usepackage{dcolumn}
\usepackage{epsfig}
\usepackage{grffile}
\usepackage[latin1]{inputenc}
\usepackage{latexsym}
\usepackage{rotating}
\usepackage{hyperref}
\usepackage[usenames]{color}
\usepackage{float}
\usepackage{xspace} 
\usepackage{mathrsfs}
\usepackage{subfigure}
\usepackage{multirow}
\usepackage{booktabs}
\usepackage{soul}
\usepackage{lipsum}% http://ctan.org/pkg/lipsum
\usepackage{graphicx}% http://ctan.org/pkg/graphicx
\usepackage{colortbl}
\usepackage{etoolbox}
\usepackage[table]{xcolor}

\def\prd{{\it Phys. Rev.} D~}

\newcolumntype{C}[1]{>{\centering\arraybackslash}p{#1}}

\def\lesssim{\mathrel{\hbox{\rlap{\hbox{\lower4pt\hbox{$\sim$}}}\hbox{$<$}}}}
\def\gtrsim{\mathrel{\hbox{\rlap{\hbox{\lower4pt\hbox{$\sim$}}}\hbox{$>$}}}}
\def\alt{\mathrel{\hbox{\rlap{\hbox{\lower4pt\hbox{$\sim$}}}\hbox{$<$}}}}
\def\agt{\mathrel{\hbox{\rlap{\hbox{\lower4pt\hbox{$\sim$}}}\hbox{$>$}}}}

\usepackage{mathtools}

\newenvironment{cititemize2}
{\begin{list}{$\bullet$}
        {\setlength{\topsep}{0pt}
         \setlength{\itemsep}{0pt}
         \setlength{\parsep}{0.25\parsep}
         \settowidth{\labelwidth}{$\bullet$}
         \setlength{\leftmargin}{1em}
}
}
{\end{list}}

\def\gta{\ifmmode {\mathbin{\lower 3pt\hbox   %> or of order
    {$\,\rlap{\raise 5pt\hbox{$\char'076$}}\mathchar"7218\,$}}}
    \else {${\mathbin{\lower 3pt\hbox
    {$\rlap{\raise 5pt\hbox{$\char'076$}}\mathchar"7218\,$}}}
    $}\fi}
\def\lta{\ifmmode {\,\mathbin{\lower 3pt\hbox   %< or of order
    {$\,\rlap{\raise 5pt\hbox{$\char'074$}}\mathchar"7218\,$}}}
    \else {${\mathbin{\lower 3pt\hbox
    {$\rlap{\raise 5pt\hbox{$\char'074$}}\mathchar"7218\,$}}}
    $}\fi}

\newcommand{\msun}{{\rm M}_{\odot}}
\newcommand{\beq}{\begin{equation}}
\newcommand{\eeq}{\end{equation}}
\newcommand{\bea}{\begin{eqnarray}}
\newcommand{\eea}{\end{eqnarray}}

\definecolor{darkperiwinkle}{RGB}{102, 102, 128}

\newcommand{\NCSA}{\affiliation{National Center for Supercomputing Applications, University of Illinois at Urbana-Champaign, Urbana, Illinois 61801, USA}}
\newcommand{\CAII}{\affiliation{NCSA Center for Artificial Intelligence Innovation, University of Illinois at Urbana-Champaign, Urbana, Illinois 61801, USA}}
\newcommand{\ANCSA}{\affiliation{Department of Astronomy, University of Illinois at Urbana-Champaign, Urbana, Illinois 61801, USA}}
\newcommand{\PNCSA}{\affiliation{Department of Physics, University of Illinois at Urbana-Champaign, Urbana, Illinois 61801, USA}}

\newcommand{\Math}{\affiliation{Department of Mathematics, University of Illinois at Urbana-Champaign, Urbana, Illinois, 61801}}
\newcommand{\iCASU}{\affiliation{Illinois Center for Advanced Studies of the Universe, University of Illinois at Urbana-Champaign, Urbana, Illinois, 61801, USA}}

\definecolor{light-gray}{gray}{0.9}

\newcommand{\comment}[1]{}
\newcommand{\RNum}[1]{\uppercase\expandafter{\romannumeral #1\relax}}

\begin{document}

\title{Deep Learning Ensemble for Real-time Gravitational Wave Detection \\ of Spinning Binary Black Hole Mergers}

\author{Wei Wei}\NCSA\CAII\PNCSA
\author{Asad Khan}\NCSA\CAII\PNCSA
\author{E. A. Huerta}\NCSA\CAII\PNCSA\iCASU\ANCSA
\author{Xiaobo Huang}\NCSA\CAII\Math
\author{Minyang Tian}\NCSA\CAII\PNCSA

\date{\today}

\begin{abstract}
\noindent We introduce the use of deep learning ensembles for 
real-time, gravitational wave detection of spinning binary black hole mergers. 
This analysis consists of training independent neural networks that 
simultaneously process strain data from multiple detectors. 
The output of these networks is then combined and processed to 
identify significant noise triggers. We have applied this methodology in O2 and O3 
data finding that deep learning ensembles clearly identify binary black hole 
mergers in open source data available at the \texttt{Gravitational-Wave Open Science Center}.
We have also benchmarked the performance of this new methodology by 
processing 200 hours of open source, advanced LIGO noise from August 2017.
Our findings indicate that our approach identifies real gravitational 
wave sources in advanced LIGO data with a false positive rate of 
1 misclassification for every 2.7 days of searched data. A follow up of 
these misclassifications identified them as glitches. 
Our deep learning ensemble represents 
the first class of neural network classifiers that are 
trained with millions of modeled waveforms that describe quasi-circular, spinning, 
non-precessing, binary black hole mergers. Once fully trained, our deep learning 
ensemble processes advanced LIGO strain data faster than real-time using 4 NVIDIA V100 GPUs.
\end{abstract}

\pacs{Valid PACS appear here}% PACS, the Physics and Astronomy
\maketitle

%%%%%%%%%%%%%%%%%%%%%%%%%%%%%%%%%%%%%%%%%%%%%
%%%%%%%%%%%%%%%%%%%%%%%%%%%%%%%%%%%%%%%%%%%%%

\section{Introduction}
\label{sec:intro}

The advanced LIGO~\cite{LSC:2015}  and Virgo~\cite{Virgo:2015} detectors have 
reported over fifty 
gravitational wave observations by the end of their third observing run~\cite{Abbott:2020niy}. 
Gravitational wave detection is now routine. It is then timely and necessary to 
accelerate the development and adoption of signal processing tools that minimize 
time-to-insight, and that optimize the use of available, oversubscribed 
computational resources. 

Over the last decade, deep learning has emerged as a go-to tool to address 
computational grand challenges across disciplines. It is extensively documented that innovative 
deep learning applications in industry and technology have addressed 
big data challenges that are remarkably similar 
to those encountered in gravitational wave astrophysics. It is then worth harnessing these developments 
to help realize the science goals of gravitational wave astrophysics in the big data era. 

The use of deep learning to enable real-time gravitational wave observations 
was first introduced in~\cite{geodf:2017a} in the context of simulated advanced LIGO noise, and then 
extended to real advanced LIGO noise in~\cite{George:2017qtr,GEORGE201864}. Over the last few years this 
novel approach has been 
explored in earnest~\cite{2018GN,Skliris:2020qax,Lin:2020aps,Wang:2019zaj,Nakano:2018vay,Fan:2018vgw,Li:2017chi,Deighan:2020gtp,Miller:2019jtp,Krastev:2019koe,2020PhRvD.102f3015S,Dreissigacker:2020xfr,Khan:2020foe,Dreissigacker:2019edy,2020PhRvD.101f4009B,2020arXiv200914611S,Khan:2020fso,PhysRevLett.122.211101,wei_warning}. However, several challenges remain. To mention a few, 
deep learning detection algorithms continue to use shallow signal manifolds which typical involve 
only the masses of the binary components. There is also a pressing need to 
develop models that process long datasets in real-time while ensuring that they 
keep the number of misclassifications at a minimum. This article introduces the 
use of deep learning ensembles to address these specific issues. 

We showcase the application of this approach by identifying all binary black hole mergers 
reported during advanced LIGO's second and third observing runs. We also demonstrate that 
when we feed 200 hours of advanced LIGO data into our deep learning ensemble, this method 
is capable of clearly identifying real events, while also significantly reducing the number of 
misclassifications to just 1 for every 2.7 days of searched data. When we followed up these 
misclassifications, we realized that they were loud glitches in Livingston data.

\subsection{Executive Summary}

At a glance the main results of this article are:

\begin{cititemize2}
\item We introduce the first class of neural network classifiers that sample a 4-D signal manifold 
that describes quasi-circular, spinning, non-precessing binary black hole mergers.
\item We use deep learning ensembles to search for and detect real binary 
black hole mergers in open source data from the second and third observing runs available at the 
\texttt{Gravitational-Wave Open Science Center}~\cite{Vallisneri:2014vxa}.
\item Our deep learning ensemble is used to process 200 hours of open source advanced LIGO noise, 
finding that this methodology: (i) processes data faster than real-time using a single GPU; (ii) clearly 
identifies real events in 
this benchmark dataset; and (iii) reports only three misclassifications that are associated 
with loud glitches in Livingstone data. 
\end{cititemize2}

This paper is organized as follows. Section~\ref{sec:architecture} describes the architecture of the 
neural networks used to construct the ensemble. Section~\ref{sec:data} summarizes the modeled waveforms 
and real advanced LIGO noise used to train the ensemble. We summarize our results in  
 Section~\ref{sec:results}, and outline future directions of research in Section~\ref{sec:end}.

\section{Neural Network Architecture}
\label{sec:architecture}

It is now well established that the design of a neural network architecture is as 
critical as the choice of optimization schemes~\cite{pidl1,pidl2}. Based on previous studies we 
have conducted to denoise real gravitational wave signals~\cite{Wei:2019zlc}, 
and to characterize the 
signal manifold of quasi-circular, spinning, non-precessing, binary black hole mergers~\cite{dlhubmodel1,Khan:2020foe}, 
we have selected \texttt{WaveNet}~\cite{2016wavenet} as the baseline architecture for 
gravitational wave detection. As we describe below, we have modified the original architecture 
with a number of important features tailored for signal detection.

\subsection{Primary Architecture}
\texttt{WaveNet}~\cite{2016wavenet} has been 
extensively used to process waveform-type time-series data, such as raw audio waveforms 
that mimic human speech with high fidelity. It is known to adapt well to time-series of high 
sample rate, as its dilated convolution layers allows larger reception fields with fewer 
parameters, and its blocked structure allows response to a combination of frequencies ranges. 

Since we are using 
\texttt{WaveNet} for classification instead of waveform generation, we have removed the 
causal structure of the network described in~\cite{2016wavenet}. The causal structure 
of \texttt{WaveNet} is modeled with a convolutional layer~\cite{krizhevsky2012imagenet} 
with kernel size 2, and by shifting the output of a normal convolution by a few time steps. 
However, in this paper we adopt convolutional layers with kernel size 3, so that the neural 
network will take into consideration both past and future information when deciding on the 
label at the current time step. We also dilate the convolutional layers to get an exponential 
increase in the size of the receptive field~\cite{oord2016wavenet}. This is necessary to 
capture long-range correlations, as well as to increase computational efficiency.  By 
construction, \texttt{WaveNet} utilizes deep residual learning, which is specifically 
tailored to train deeper neural network models~\cite{he2016deep}. The structure of 
\texttt{WaveNet} is described in detail in~\cite{oord2016wavenet}. Below we describe 
tailored, \texttt{WaveNet}-based architectures for gravitational wave detection.

\subsection{\texttt{WaveNet} Architecture I}
\label{architecture_I}

\begin{figure}
\centerline{
\includegraphics[width=0.5\textwidth]{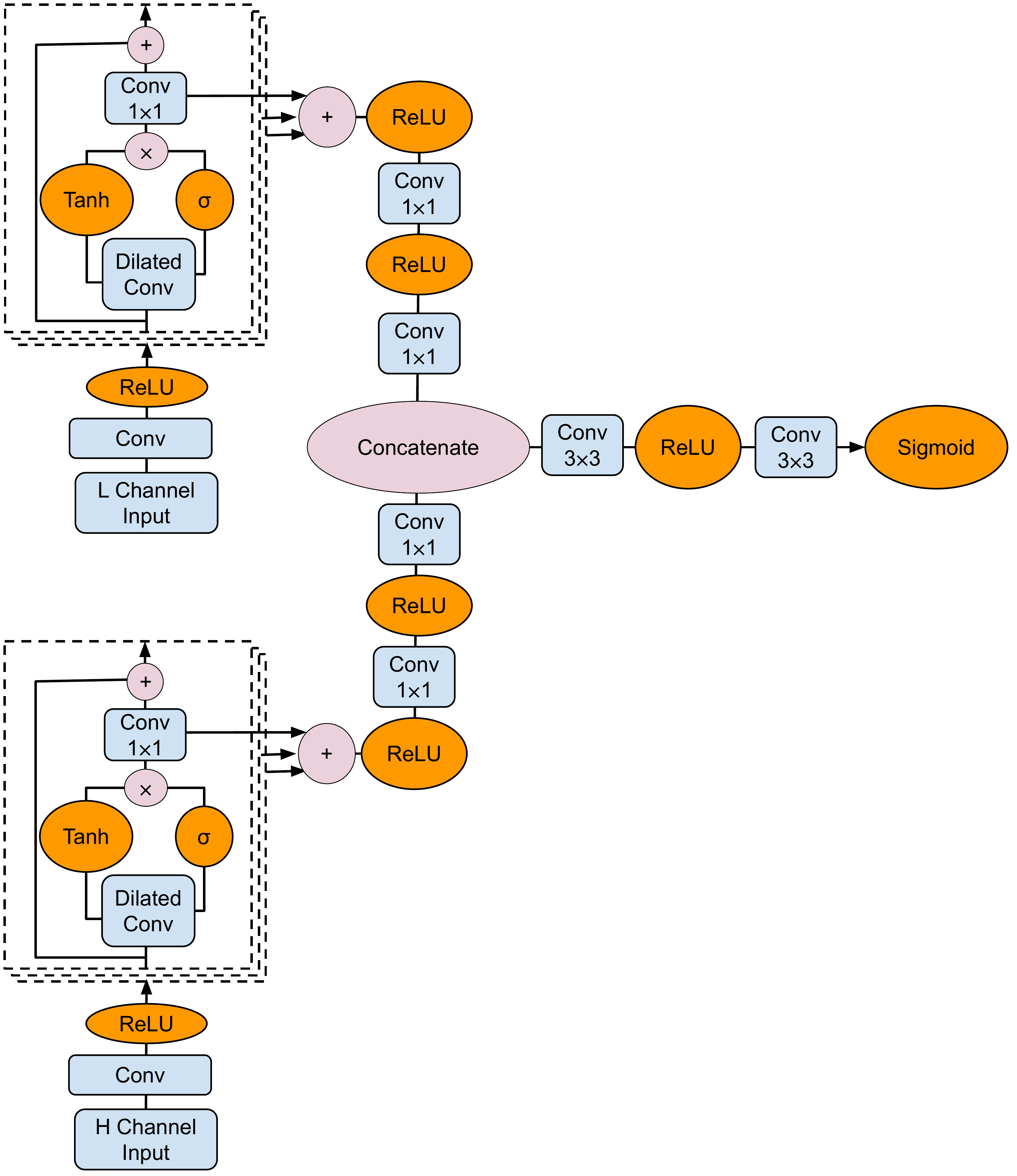}
} 
\caption{Architecture of our \texttt{WaveNet} detection algorithm. The input and 
output are tensors of shape $\textrm{batch\_size}\times1\times\texttt{model}$, 
where \(\texttt{model}=\{16384,\,4096\}\).}
\label{fig:net}
\end{figure}

Model \RNum{1} processes Livingston and Hanford strain data using two independent 
\texttt{WaveNet} models. Then the output of the two \texttt{WaveNet}s is concatenated, 
and fed into the last two convolutional layers. Finally, a \texttt{sigmoid} transformation is 
applied to ensure that the output values are in the range of $[0,1]$. The network 
structure is shown in Figure~\ref{fig:net}.

\subsection{\texttt{WaveNet} Architecture II}
\label{architecture_II}

Model \RNum{2} is essentially the same shown in Figure~\ref{architecture_I}, except 
that the input now consists of $1$s long strain data sampled at $4096$ Hz. 
Additionally, we reduce the depth of the model by cutting down to $4$ residual blocks, 
and reduce the number of filters, and revert back to kernel size $2$.

\section{Data Curation}
\label{sec:data}

In this section we describe the modeled waveforms used for training, 
and the strategy followed to combine these signals with 
real advanced LIGO noise.

\subsection{Modeled Waveforms}

\begin{figure}
\includegraphics[width=0.5\textwidth]{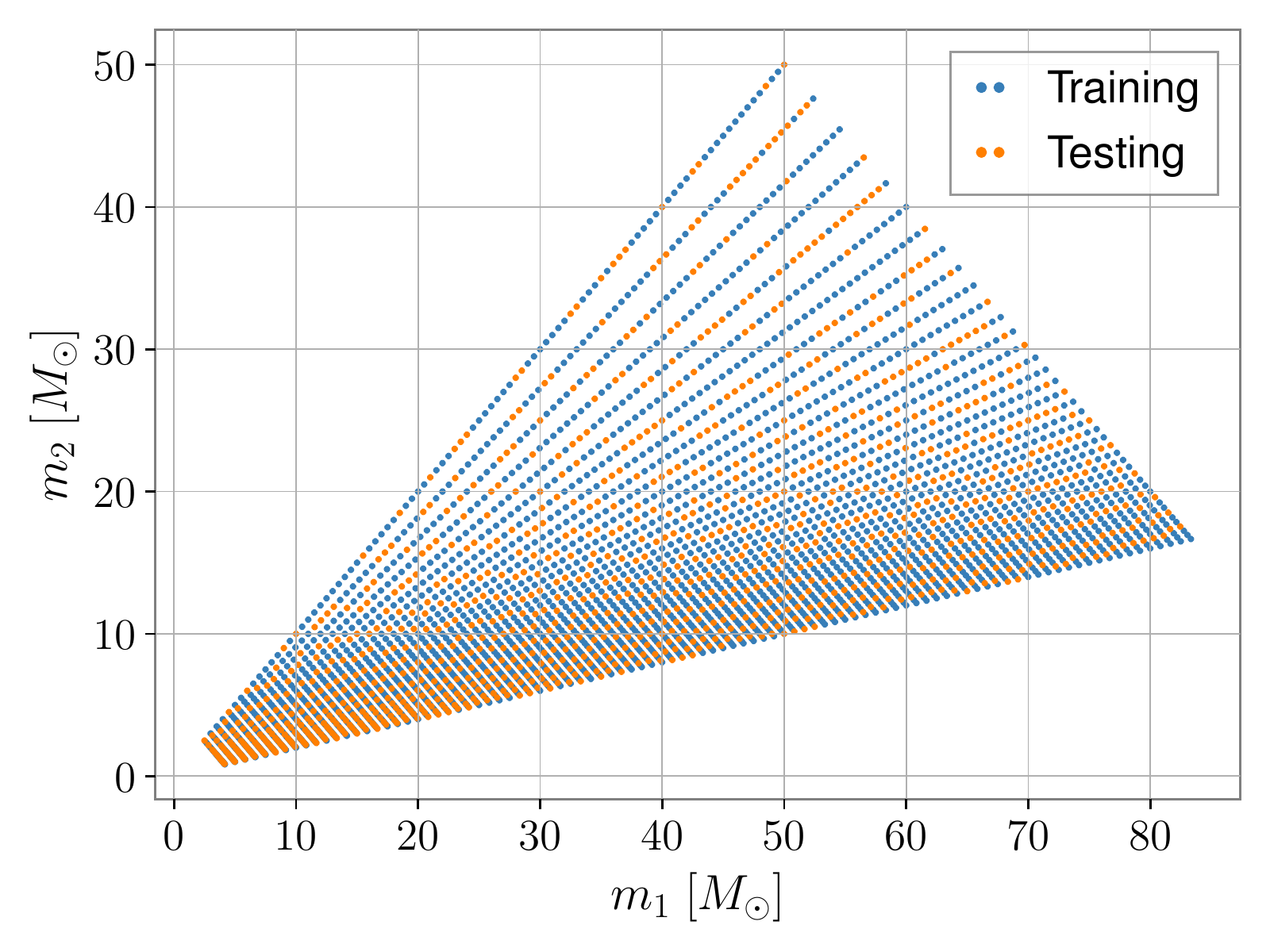}
\includegraphics[width=0.5\textwidth]{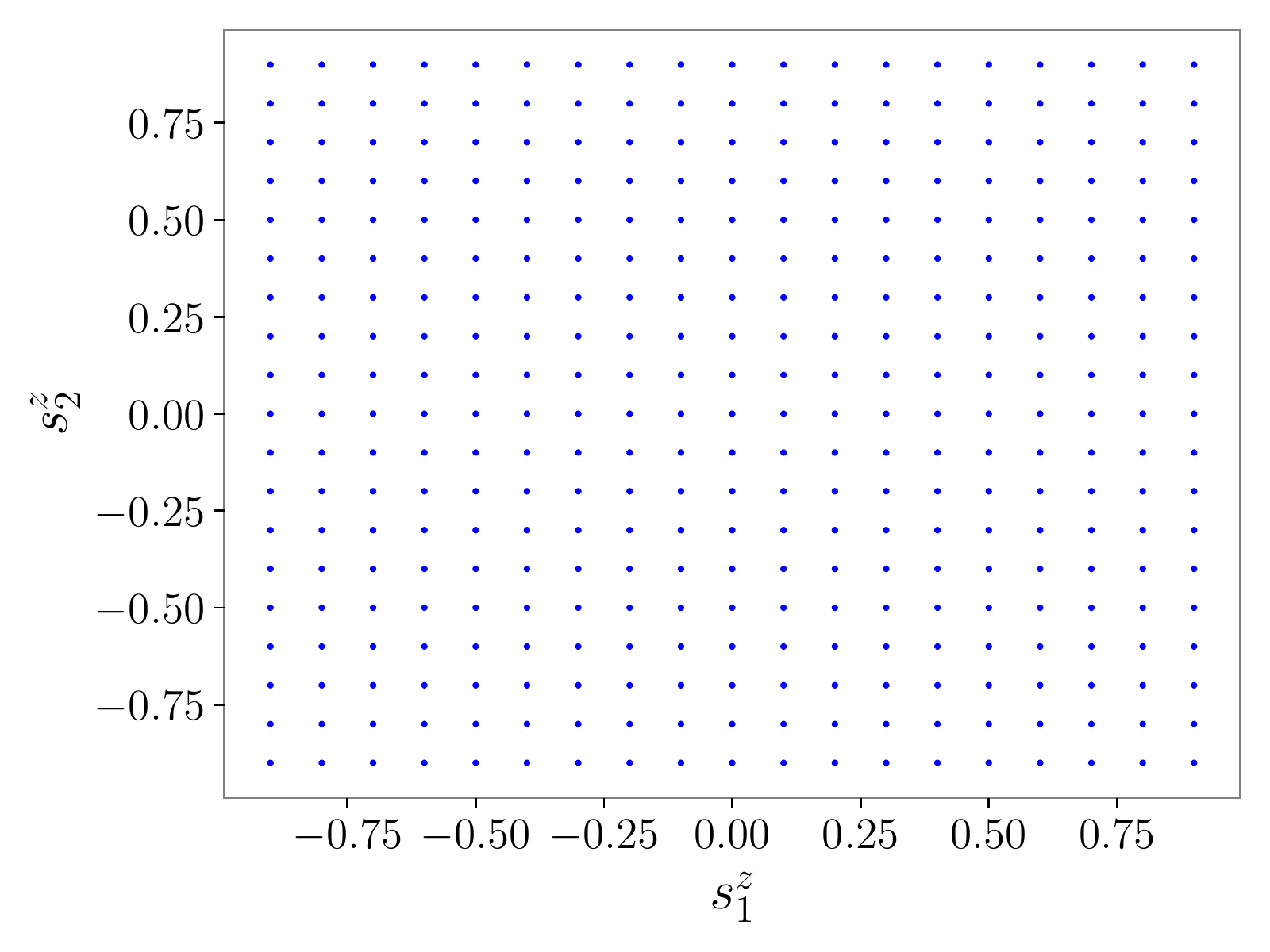}
\caption{Sampling of the component mass and individual spin parameter space 
\((m_1,\,m_2,\, s_1^z,\,s_2^z)\) for training and testing.}
\label{fig:coverage}
\end{figure}

We train our neural networks using \texttt{SEOBNRv3} waveforms~\cite{seobnrv3}. 
Our datasets consist of time-series waveforms that describe the last second of evolution 
that includes the late inspiral, 
merger and ringdown of quasi-circular, spinning, non-precessing, binary black hole mergers. 
Each waveform is produced at a sample rate of 
16384Hz and 4096Hz. The parameter space covered for training encompasses total masses 
\(M\in[5\msun,\,100\msun]\), mass-ratios \(q \leq 5\), and individual spins \(s^z_{\{1,2\}}\in[-0.8,\,0.8]\). 
The sampling of this 4-D parameter space is shown in Figure~\ref{fig:coverage}. 
It is worth pointing out that even though these models cover a total mass range \(M\leq100\msun\), our models 
are able to generalize, since they can clearly identify the O3 event GW190521, which has an 
estimated total mass \(M\sim142\msun\)~\cite{Abbott:2020tfl}.

We consistently encode ground truth labels for waveforms in a binary manner, 
where data points before the amplitude peak of every waveform are labeled 
as 1, while the data points after the merger are labelled as 0. 
In other words, the change from 1's to 0's indicates the location of the merger. 

\subsection{Advanced LIGO noise for training}
\label{sec:preprocessing}

We prepare the noise used for training by selecting continuous segments of advanced LIGO 
noise from the \texttt{Gravitational Wave Open Science Center}~\cite{Vallisneri:2014vxa}, 
which are typically 4096 seconds long. None of these segments include known 
gravitational wave detections. These data are used to compute noise power spectral density (PSD) 
estimates~\cite{2016CQGra..33u5004U} that are used to whiten both the strain data 
and the modeled waveforms. Thereafter, the whitened strain data and the whitened 
modeled waveforms are linearly combined, and a broad range of signal-to-noise ratios 
are covered to encode scale invariance in the neural network. We then normalize the standard 
deviation of training data that contain both signals and noise to one.

We also encode time-invariance into our training data, which is critical to correctly detect 
signals in arbitrarily long data stream irrespective of their locations. For every one second 
long segments with injections used for training, the injected waveform is located at 
a random location, with the only constraint that its peak must locate inside the second 
half of the 1s-long input time series. To improve the robustness of the trained model, 
only 40\% of the samples in the training set contain GW signals, while the rest 60\% samples 
are advanced LIGO noise only.

We present a more detailed description of the training approach for Model \RNum{1} 
and Model \RNum{2} below.

\subsubsection{Training for Model \RNum{1}}

Model \RNum{1} is trained on three 4096s-long data segments with starting GPS time 
$1186725888$, $1187151872$, and $1187569664$. After we whitened the three segments 
separately with the corresponding PSD, we truncated 122s-long data from each end of 
the segment to remove edge effects. The neural network trained this way is able to 
successfully detect the gravitational wave events GW170104, GW170729, GW170809, GW170814, 
GW170817, GW170818, GW170823, GW190412. 

This neural network is also able to 
detected GW170608 and GW190521 after further trained on advanced LIGO data close to 
these two events. Specifically, to detect GW170608, the neural network is further trained on 
two 4096s-long data segments with starting GPS time $1178181632$, and 
$1180983296$. Similarly, to detect GW190521, the neural network is further 
trained on the 2048s-long data segments around GW190521.

The input to the neural network is a 1s-long $16384$Hz data segment, 
with two channels from Livingston and Hanford observatories. The output 
of the neural network has one channel and is of the same length as the input. 
The output is $1$ when there is signal at the corresponding location in the input, 
and $0$ otherwise. 

To ensure that signals contaminated by advanced LIGO noise 
generate long enough responses, we make sure that the flat peak of the 
neural network output is located in the second half of 
the 1s-long input. In other words, when there is 
a signal in the second half of the 1s-long input, the ground-truth output would be 
a peak of width of at least $8192$. Furthermore, to ensure that all possible signal 
peaks appear in the second half of the input data segment, we feed the test 
data into the neural network with a step size of $8192$, i.e., we crop out 1s-long 
data segment of size $16384$ every $8192$ data points.

To constrain the output of the neural network in the range 
$[0,1]$, we apply the sigmoid function $s(x) = 1/(1+\exp(-x))$ element-wise on 
the final output from neural network, as shown in Figure~\ref{fig:net}. We use the 
binary cross entropy loss to evaluate the prediction of the neural network when 
compared to ground-truth values. Finally, to avoid possible overfitting, we 
augment the training data by reversing the 1s-long data segments in the time 
dimension with a probability of $0.5$.

\subsubsection{Training for Model \RNum{2}}
Model \RNum{2} did not require fine-tuning on advanced LIGO data around 
any specific events, i.e., it was able to detect all O2 and O3 events after the initial round 
of training which, as mentioned above, consisted of three 4096s-long data segments 
with starting GPS time 
$1186725888$, $1187151872$, and $1187569664$. Furthermore, data augmentation 
of training data by randomly reversing the 1s long input strains was not employed.

\begin{figure*}[!htb]
\centerline{
    \includegraphics[width=0.5\textwidth]{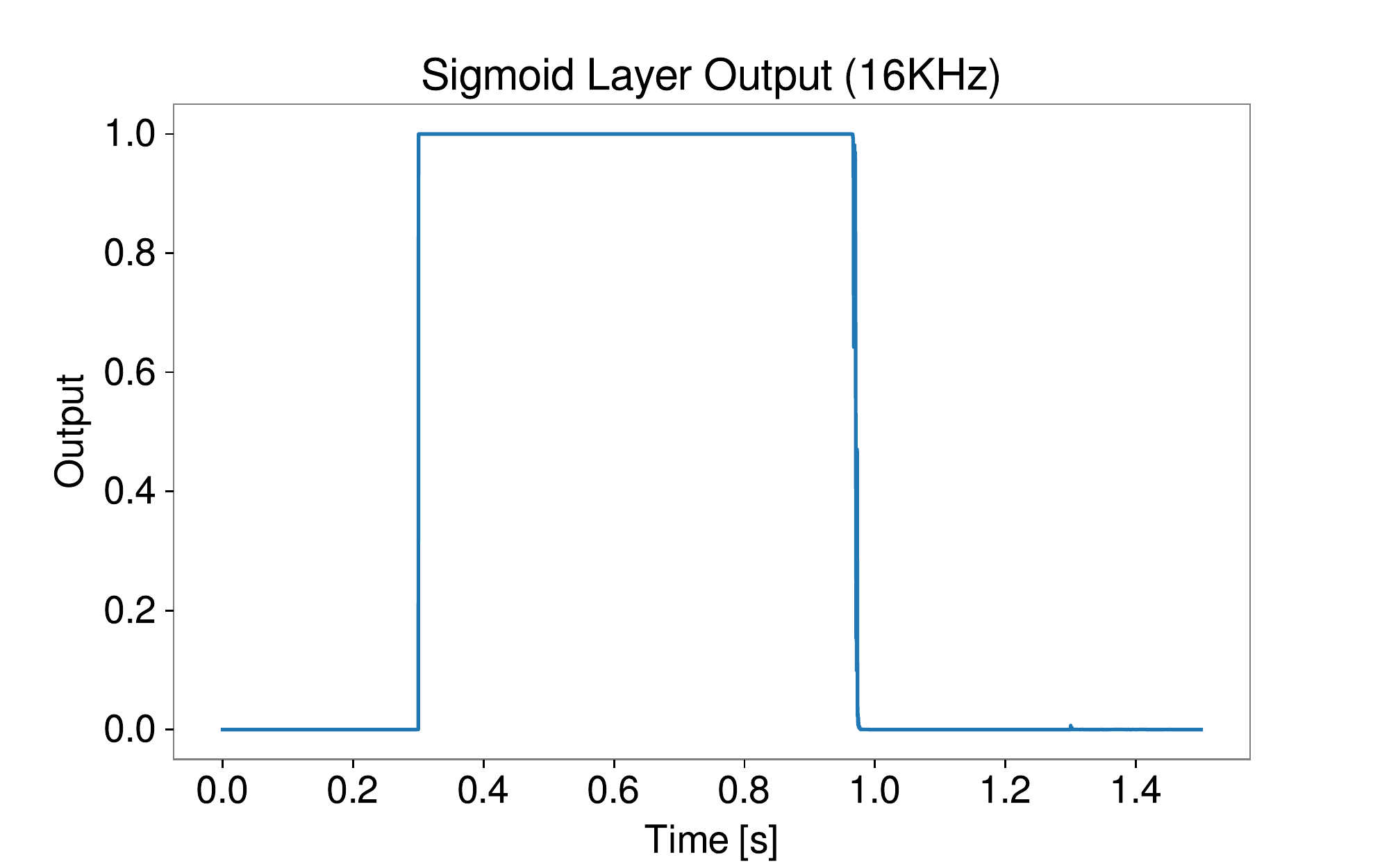}
    \includegraphics[width=0.5\textwidth]{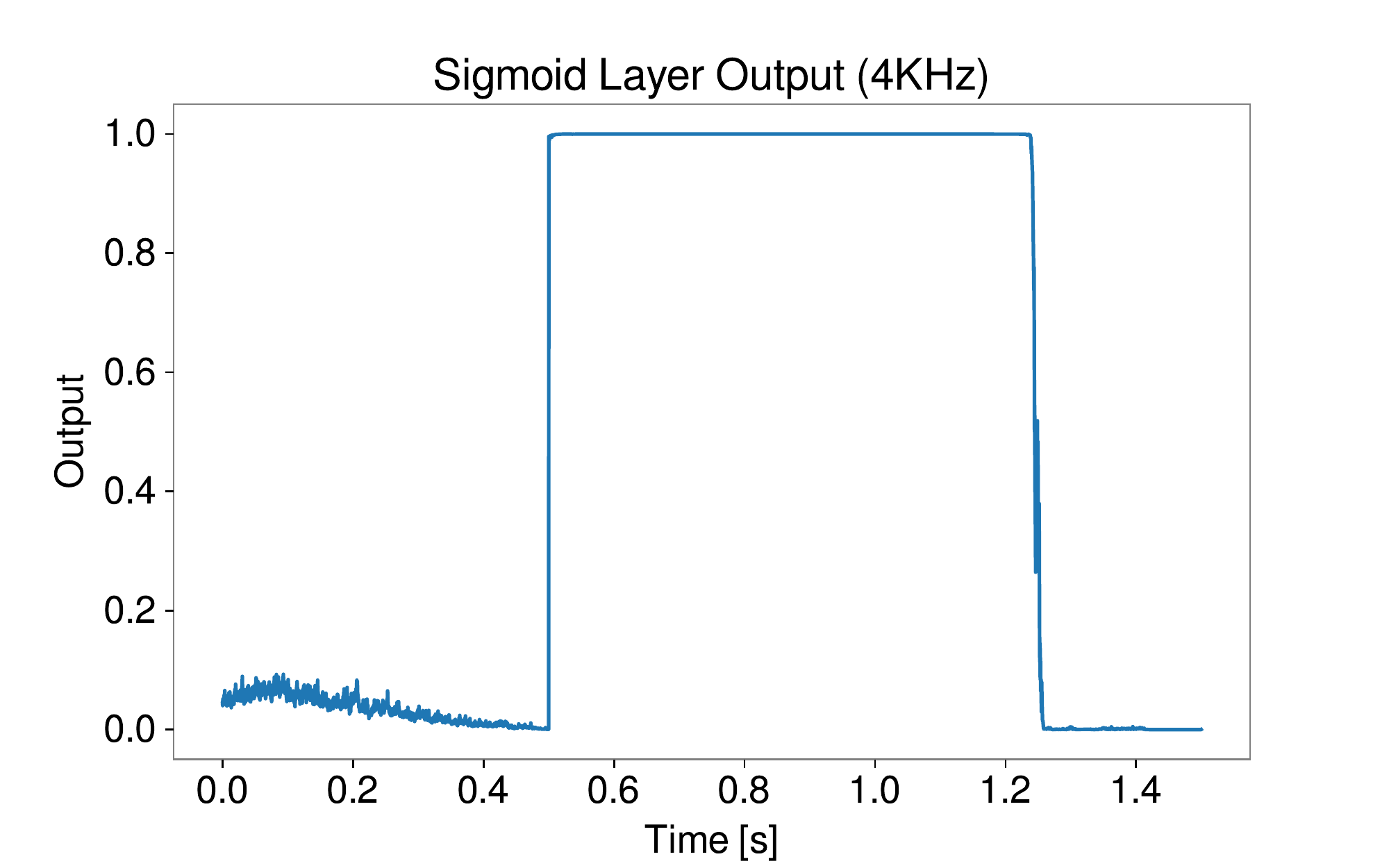}
   }
   \caption{Post-processing output of Model \RNum{1}, left panel,  and Model  \RNum{2}, 
   right panel, for the event GW170809. Notice that we have zoomed in to show the neural 
   network response in the vicinity of the waveform signal.}
   \label{fig:output}
\end{figure*}

\subsection{Optimization Methodology}

\textbf{Model \RNum{1}}
The neural networks are trained on 4 NVIDIA K80 GPUs in the Bridges-AI system~\cite{bridgesai}, and also on 
4 NVIDIA V100 GPUs in the Hardware Accelerated Learning (HAL) deep learning cluster~\cite{halcluster}, 
with \texttt{PyTorch}~\cite{paszke2017automatic}. We use ADAM ~\cite{kingma2014adam} optimization, 
and binary cross entropy as the loss function. The weight parameters are initialized randomly. The learning rate is 
set to $10^{-4}$.

\textbf{Model \RNum{2}}
Model \RNum{2} is trained on 8 NVIDIA V100 GPUs in the HAL cluster~\cite{halcluster} with 
\texttt{Tensorflow}~\cite{TensorFlow}, once again using ADAM~\cite{kingma2014adam} optimizer with binary cross entropy as the loss function. Similarly, the weights are initialized randomly, the initial learning rate is set to $10^{-4}$, and a step-wise learning rate scheduler is employed to attempt a more fine-grained convergence to a minima of the loss function.

\subsection{Post-processing}

Once the models process advanced LIGO strain data, their outputs are post-processed as follows. 

\subsubsection{Post-processing for Model \RNum{1}}

In the training set, for all input samples with gravitational waves, their labels will always 
have 1's before the merger and 0's after the merger, and because we always place the 
merger in the second half of the input time series, the length of the 1's will be at least 8192. 
Therefore, if the input 1s-long data strain contains a gravitational wave, the output of the neural 
network would be a peak with a height  of 1 and a width of at least $8192$. Based on this 
setup, the output of the neural network will be further fed into the off-the-shelf peak detection 
algorithm \texttt{find\_peaks} provided by \texttt{SciPy}. The algorithm will then output the 
locations of possible peaks that satisfy the conditions of at least $0.9995$ in height and 
$8192$ in width. To avoid possible overcounting, we also assume that there is at most one 
signal in a 5s-long window. Furthermore, since a GW signal will induce a flat (all 1's) and 
wide (at least 8192 in length) peak  in the neural network output, we also use an additional 
criterion that 94\% of the outputs between the left and right boundaries of the detected peak 
be greater than 0.99 to further reduce the false alarms. To showcase the application of this approach 
for real events, the left panel of Figure~\ref{fig:output} presents the output of Model 
\RNum{1} for the event GW170809.

\subsubsection{Post-processing for Model \RNum{2}}
In the post-processing, the conditions for \texttt{find\_peaks} algorithm were changed 
to $0.99993$ and $2048$ for height and width respectively. Also, the additional criterion 
was relaxed so that only 95\% of the outputs between the left and right boundaries of the 
detected peak be greater than 0.95. The right panel of Figure~\ref{fig:output} presents the 
post-processing output of Model \RNum{2} for the event GW170809. 

\subsubsection{Post-processing of deep learning ensemble}

Finally, we combine the output of the deep learning models to 
identify noise triggers that pass a given threshold of detectability 
for each model. We do this by comparing the GPS times for all triggers 
between each model. 
By definition (in the separate post-processing for each model), each model can 
only produce at most one trigger every 5 seconds. Hence, when combining the triggers 
from the two models, any triggers more than 5 seconds apart can be dropped 
as random False Alarms. In fact, since each model is extremely precise at identifying 
the merger location, we apply a much stricter criterion, namely, i.e., any triggers more 
than $1/128$ seconds apart between the two models are dropped as False Alarms, 
whereas triggers within $1/128$ seconds of each other are counted as a single 
Positive Detection.

In the following section we use this approach to search for gravitational waves in 
minutes, hours, and hundreds of hours long real advanced LIGO data.

\section{Results}
\label{sec:results}

\begin{figure*}[!htb]
\centerline{
\includegraphics[width=0.9\columnwidth]{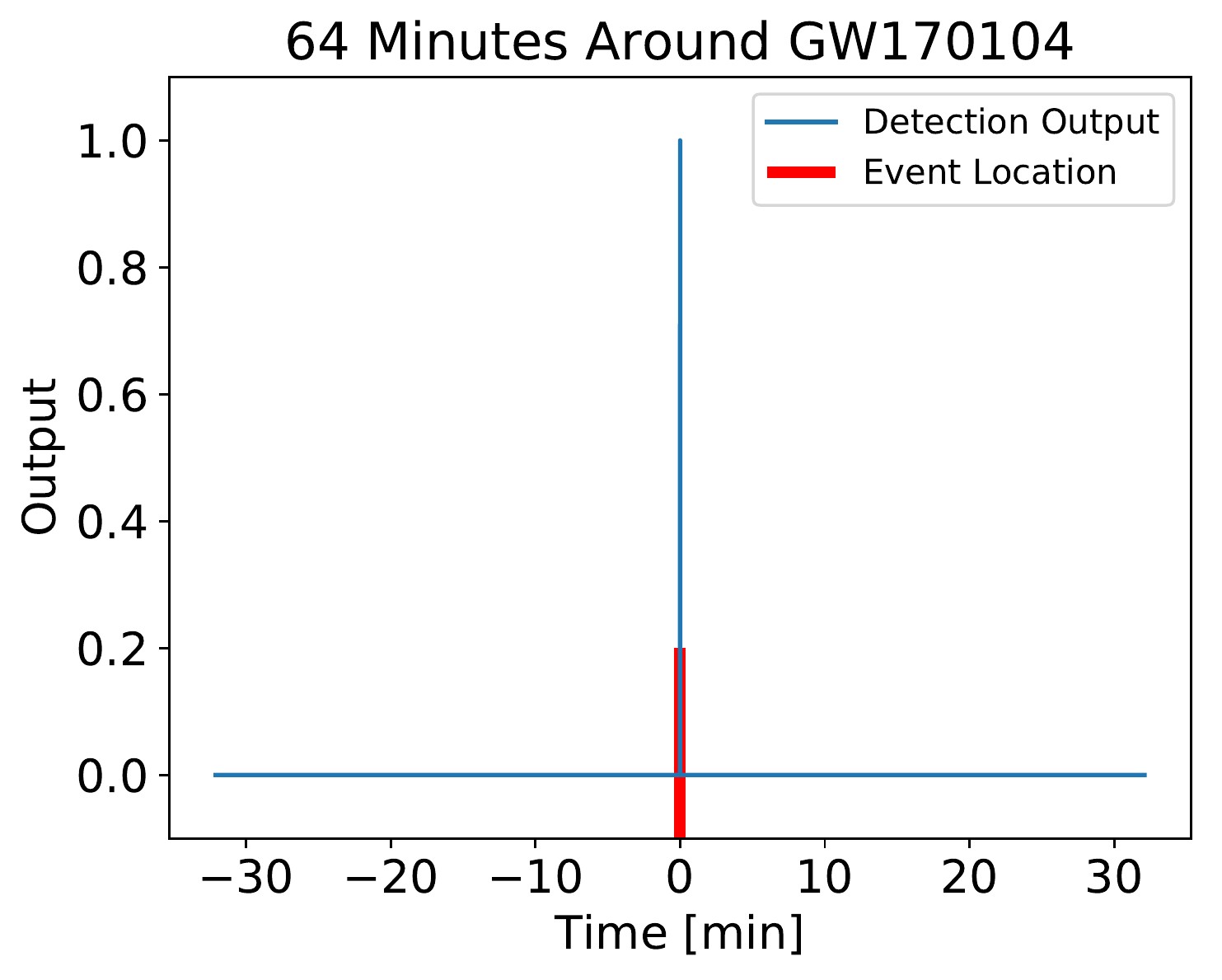}
\includegraphics[width=0.9\columnwidth]{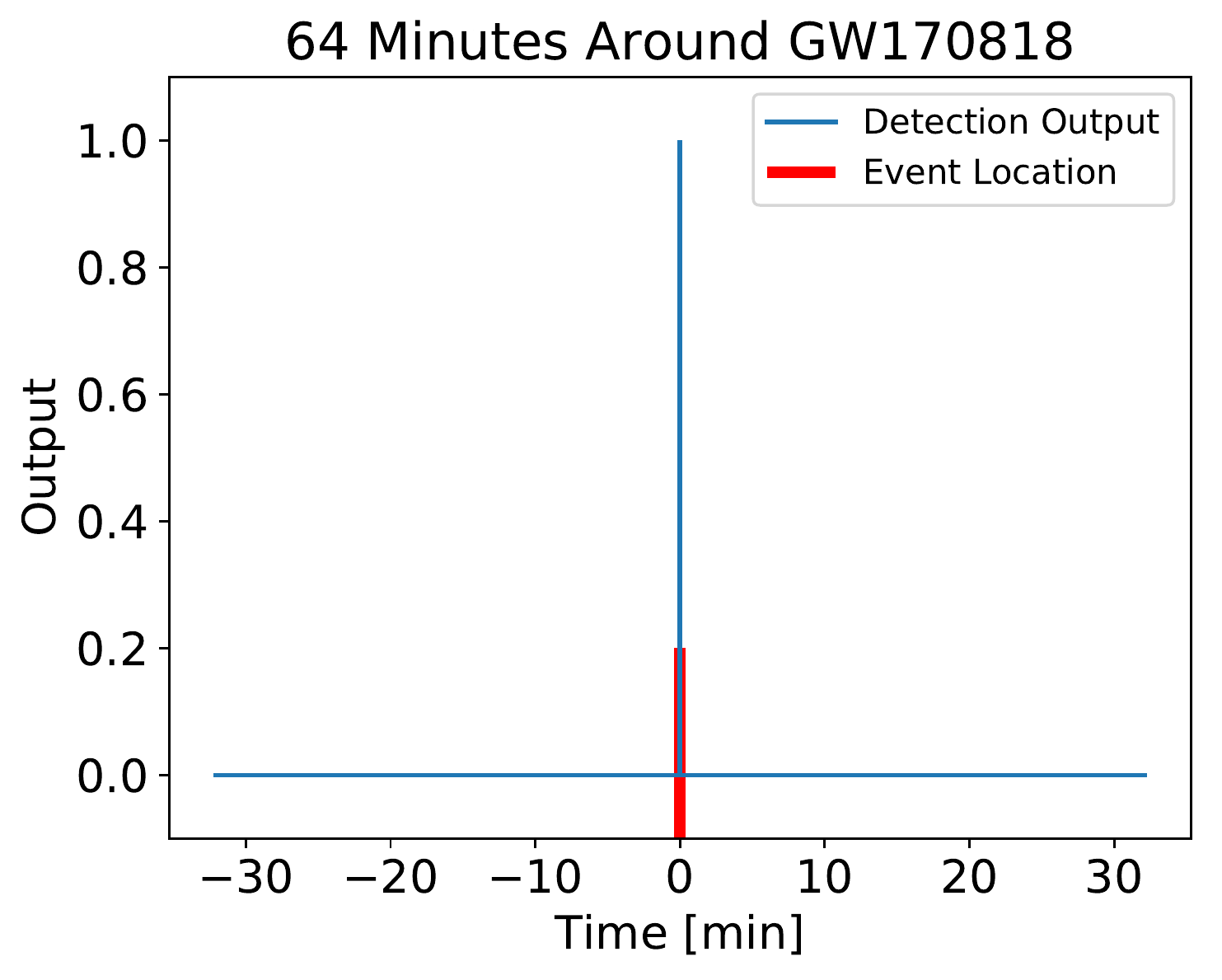}
}
\centerline{
\includegraphics[width=0.9\columnwidth]{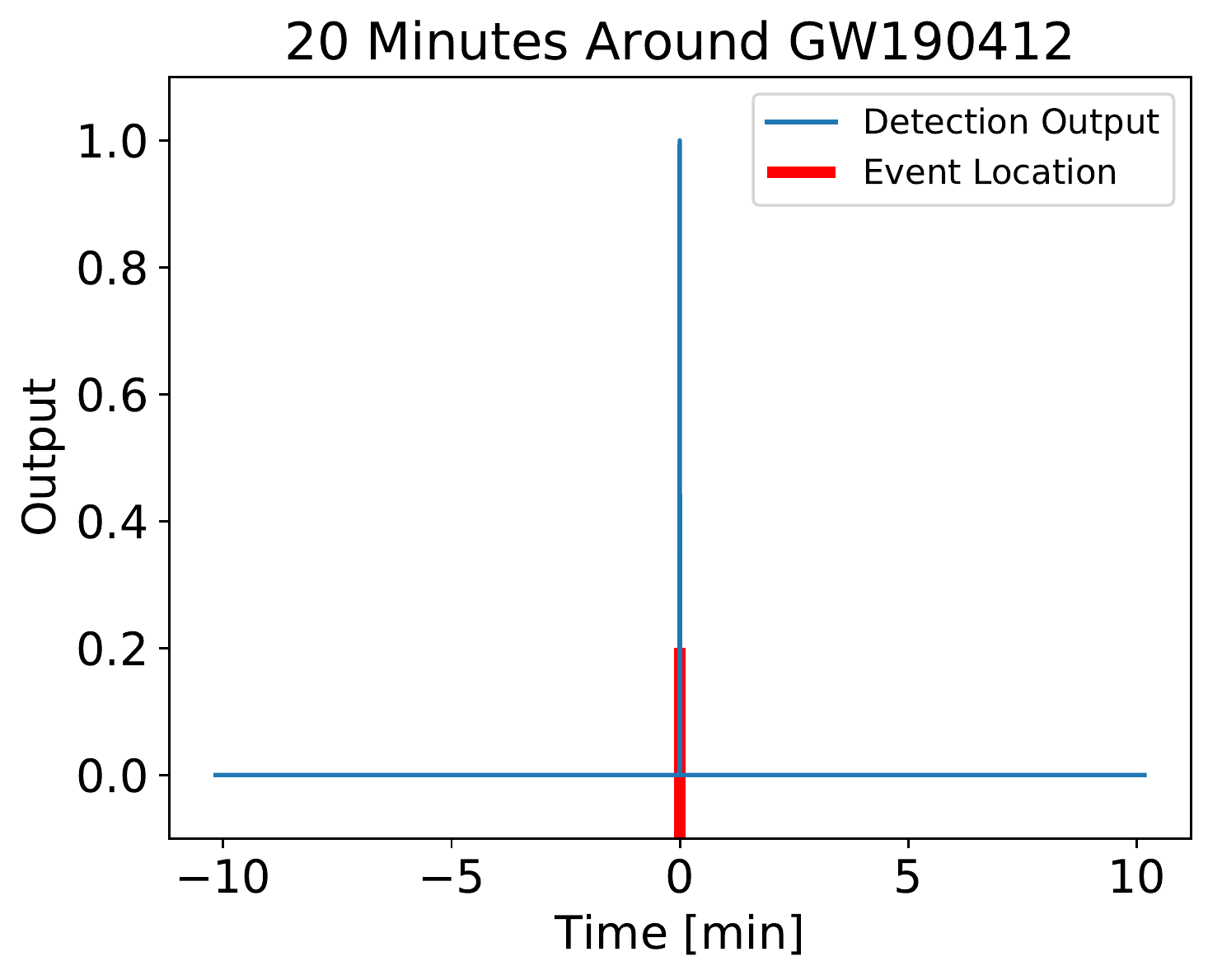}
\includegraphics[width=0.9\columnwidth]{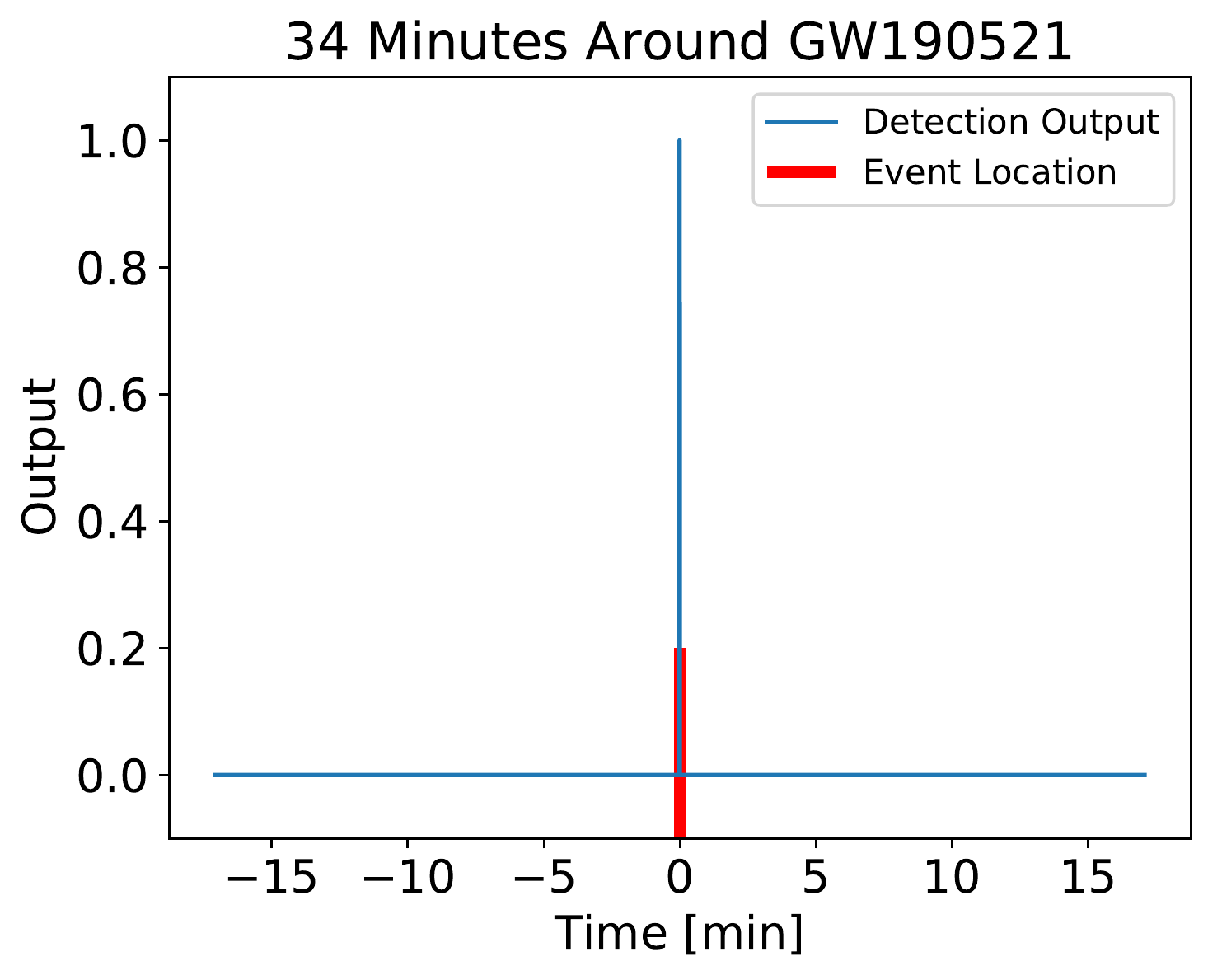}
}
\caption{Detection output of our deep learning ensemble for GW170104, top left; GW170818, top right; 
GW190412, bottom left; and GW190521, bottom right. Notice that our ensemble identifies all these 
events with no false positives in minutes and hour-long datasets.}
\label{fig:first_batch}
\end{figure*}

In this section we present results for the performance of our deep learning ensemble to 
search for and detect binary black hole mergers in O2 and O3 data. These results are summarized in 
Figures~\ref{fig:first_batch}, and Figure~\ref{fig:second_batch} in Appendix~\ref{sec:apex}. 

Figure~\ref{fig:first_batch} shows that our deep learning ensemble can detect O2 and O3 
events without any additional false positives in the minutes or hour-long datasets released 
at the \texttt{Gravitational-Wave Open Science Center} containing these events. Similar results 
may be found in Figure~\ref{fig:second_batch} in Appendix~\ref{sec:apex}. To the best of our 
knowledge this is the first time deep learning is used to search for and detect 
real events, while also reducing the number of false positives at this level, in hours-long datasets. 
Notice also that our method can generalize to detect 
events that are beyond the parameter space used to train our neural networks. This is confirmed 
by the detection of GW190521, bottom right panel in Figure~\ref{fig:first_batch}, 
which has an estimated total mass \(M\sim142\msun\), whereas 
our training dataset covered systems with total mass \(M\leq100\msun\). 

\begin{figure*}[!htb]
\centerline{
\includegraphics[width=\textwidth]{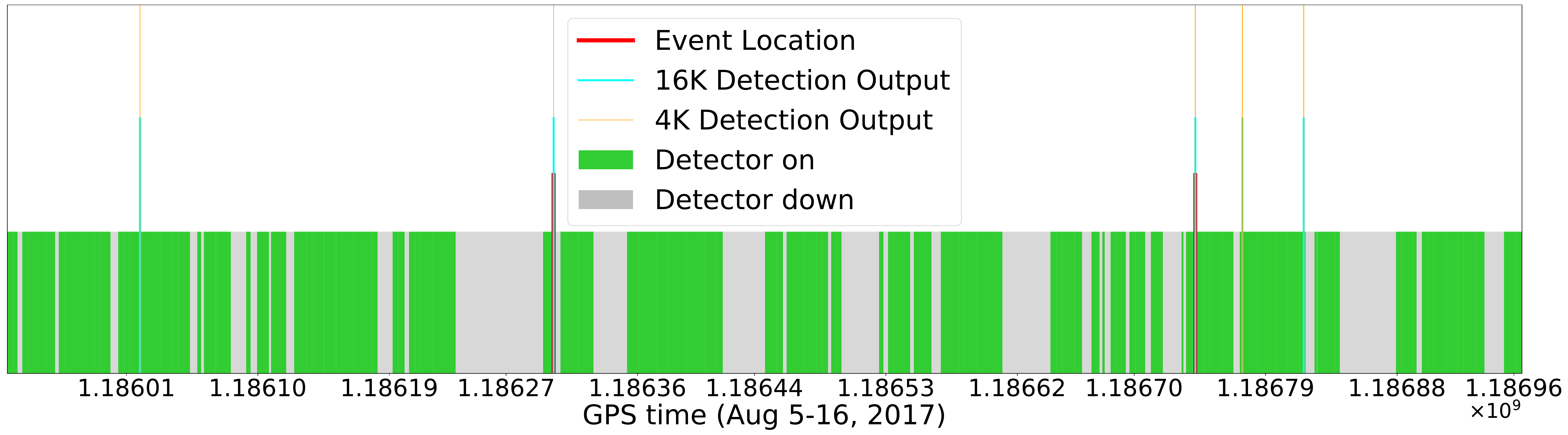}
} 
\caption{Output of our deep learning ensemble upon processing Livingston and 
Hanford data between August 5-16, 2017. This methodology identifies the two real events 
contained therein, while also indicating the existence of three false positives, associated to 
loud glitches in the Livingston channel. Every tick represents a day. These data were processed 
within 14 hours using 4 NVIDIA V100 GPUs.}
\label{fig:ensemble_long}
\end{figure*}

While this is a significant result, it is also essential to benchmark the performance of 
our approach using much longer datasets. We have done this by processing 200hrs of advanced 
LIGO noise from August 2017. We feed these data into our deep learning ensemble to address two 
issues: (i) the sensitivity of the ensemble to real events in long datasets; and (ii) quantify 
the number of false positives, and explore the nature of false positives to gain additional 
insights into the response of our deep learning ensemble to both signals and 
noise anomalies. The results of 
this analysis are presented in Figure~\ref{fig:ensemble_long}.

At a glance, Figure~\ref{fig:ensemble_long} indicates that our approach identifies two real events 
contained in this 200hr-long dataset, namely, GW170809 and GW170814. These two events are marked 
with red lines in Figure~\ref{fig:ensemble_long}. We also notice that our ensemble indicates the 
existence of three additional noise triggers, marked by blue and yellow lines, which are 
worth following up.

\begin{figure*}[!htb]
\subfigure{
    \includegraphics[width=0.52\textwidth]{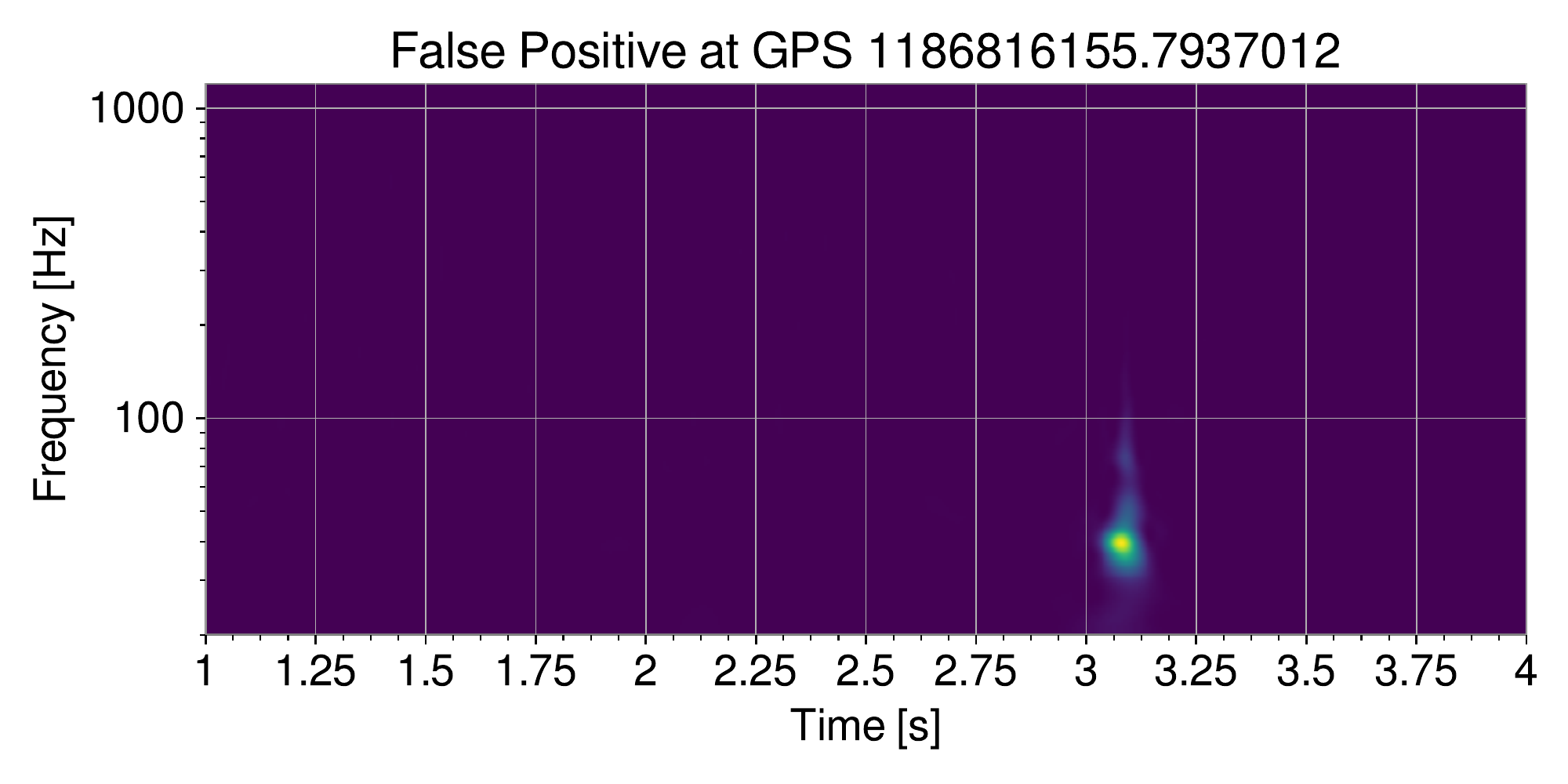}}
    \hfill
    \subfigure{\includegraphics[width=0.4\textwidth]{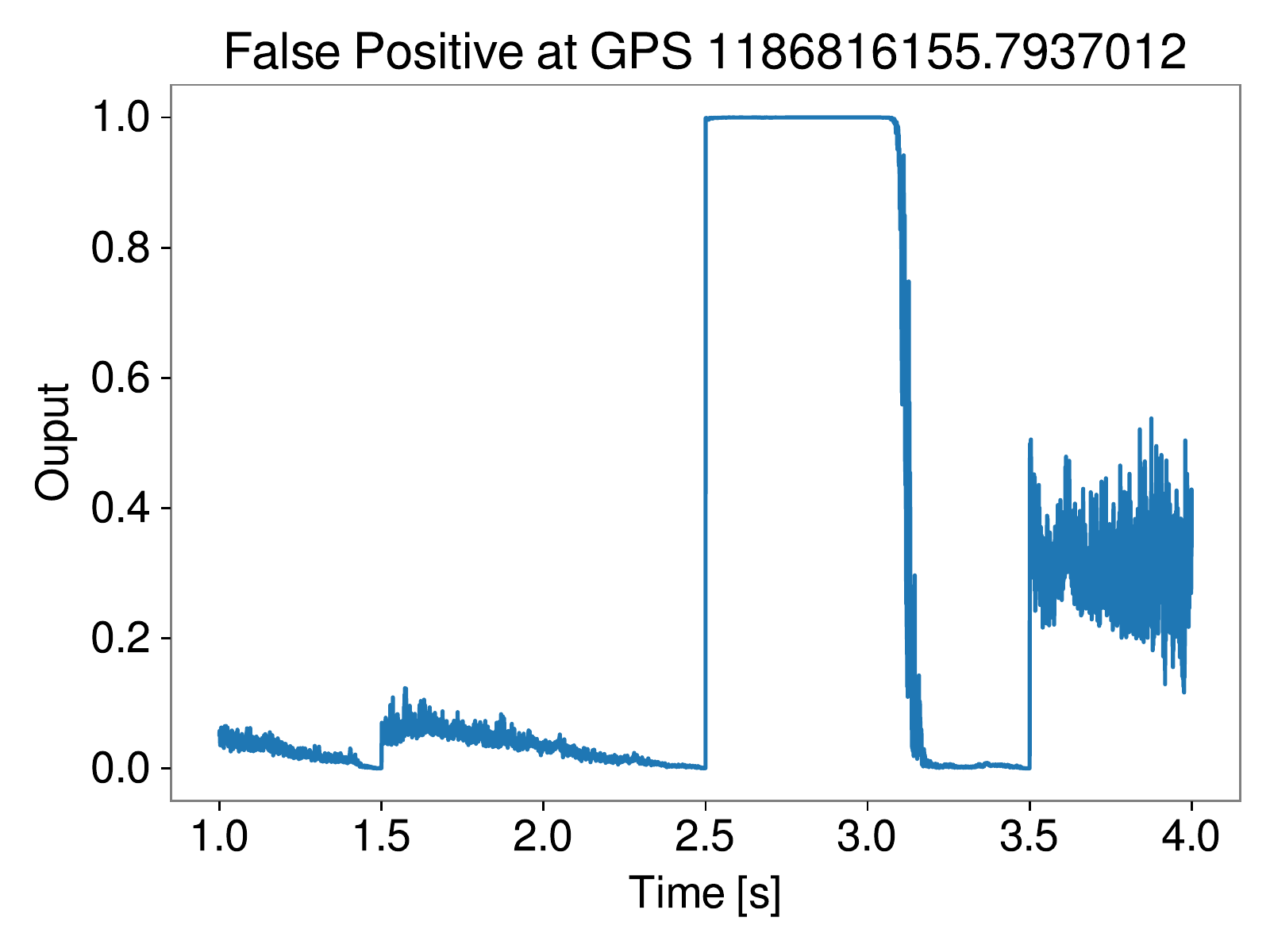}\hfill
}
 \subfigure{
    \includegraphics[width=0.52\textwidth]{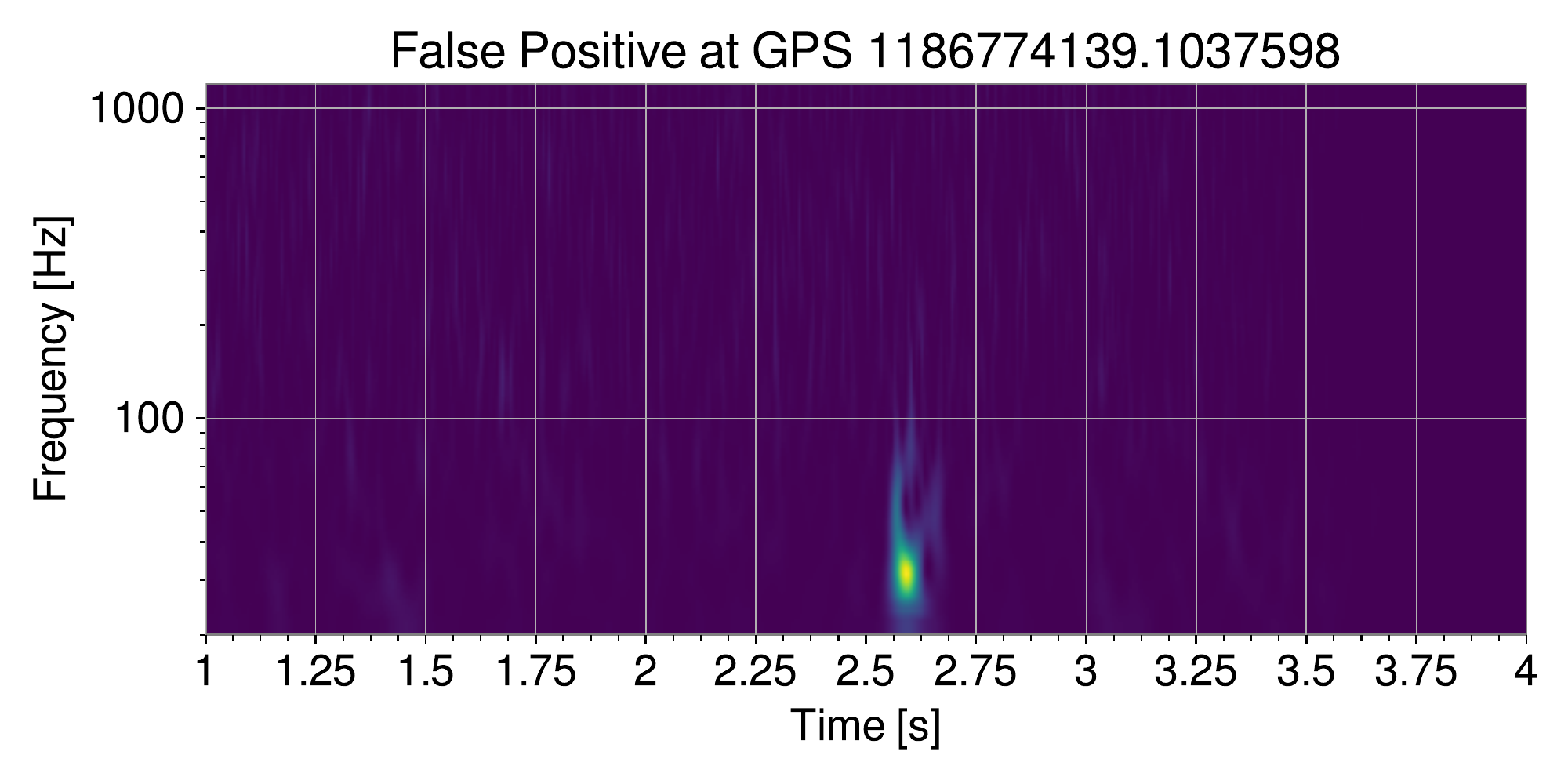}}
    \hfill
    \subfigure{\includegraphics[width=0.4\textwidth]{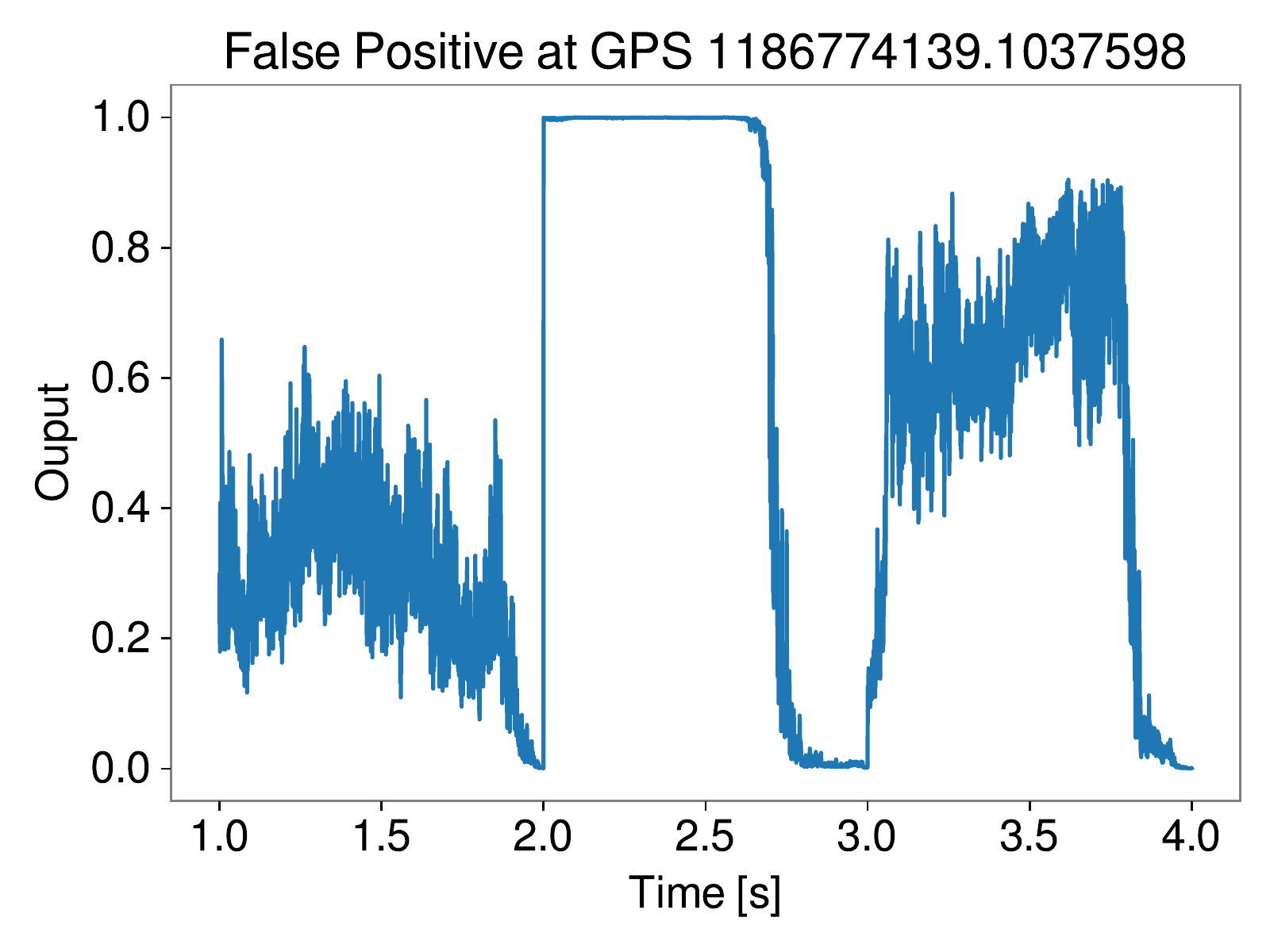}\hfill
}
 \subfigure{
    \includegraphics[width=0.52\textwidth]{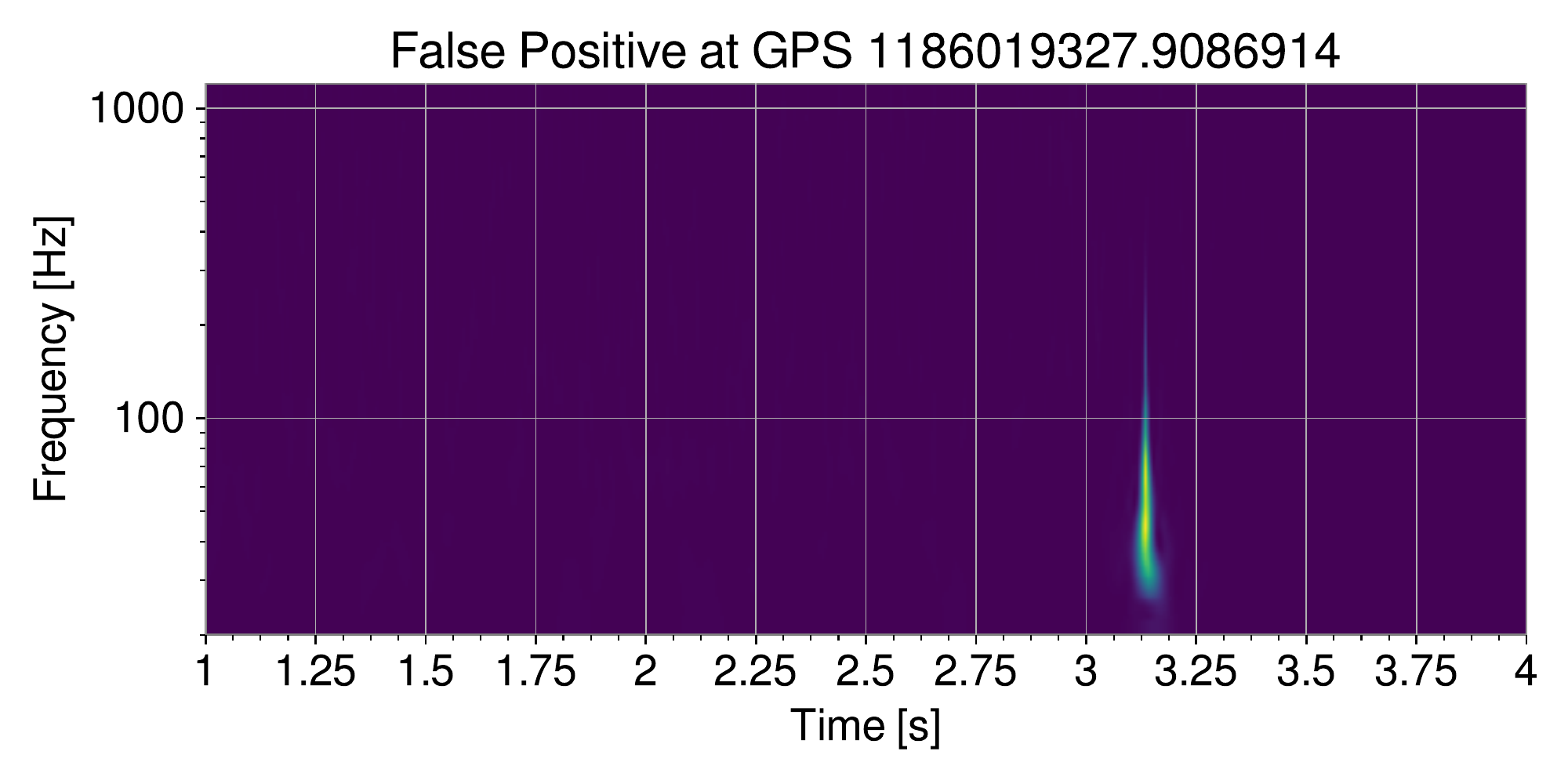}}
    \hfill
    \subfigure{\includegraphics[width=0.4\textwidth]{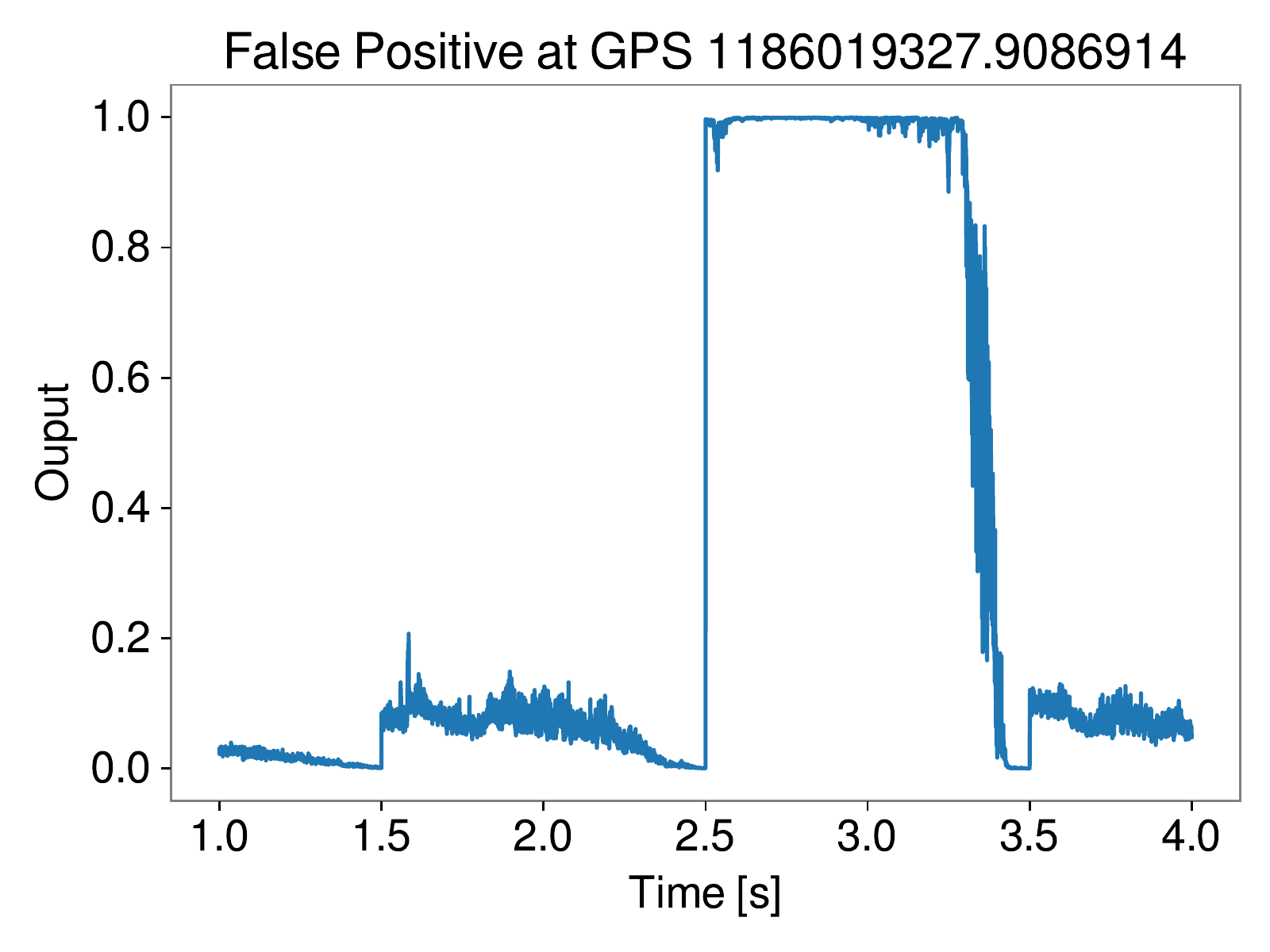}\hfill
}
 \subfigure{
    \includegraphics[width=0.52\textwidth]{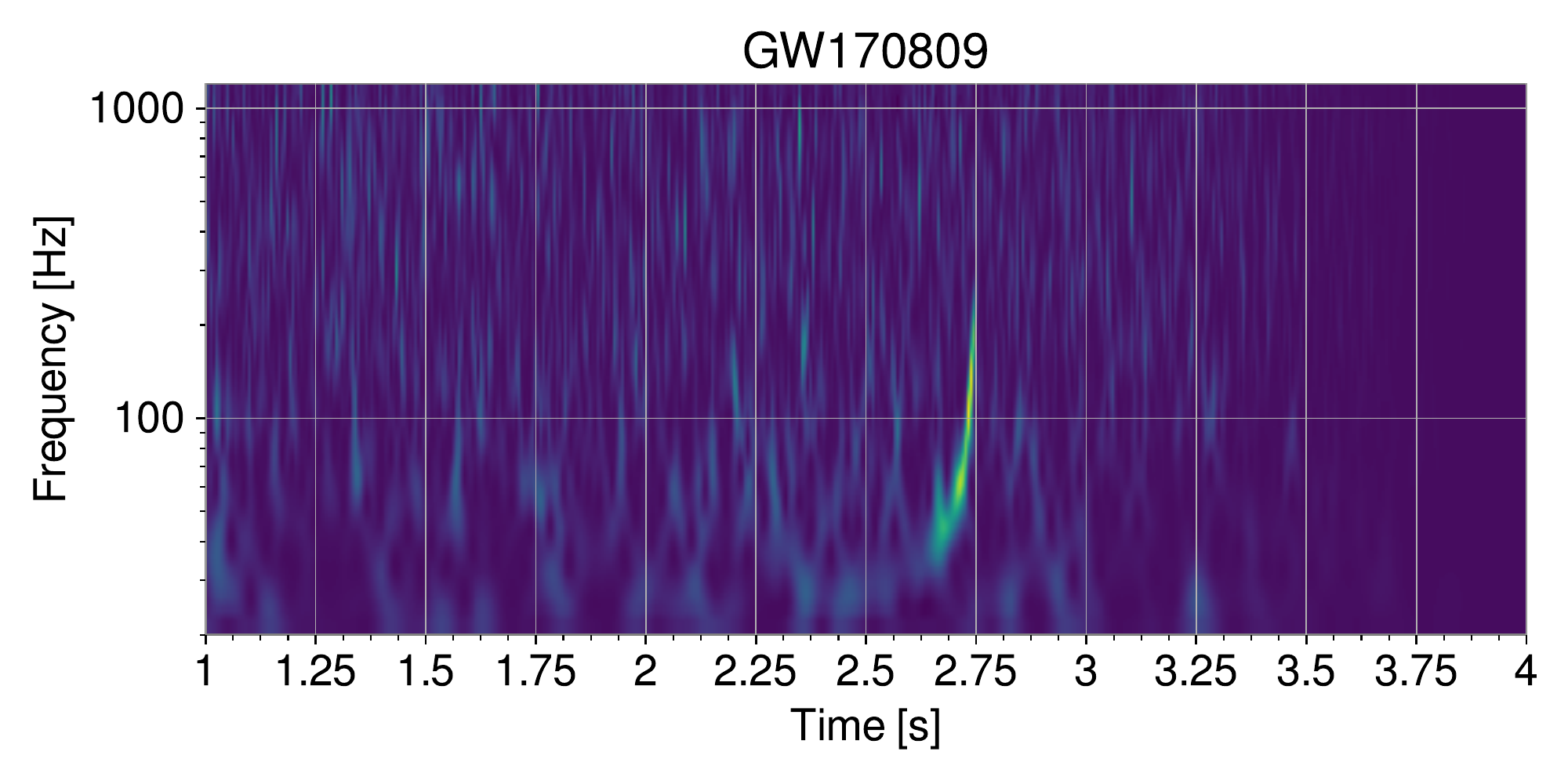}}
    \hfill
    \subfigure{\includegraphics[width=0.4\textwidth]{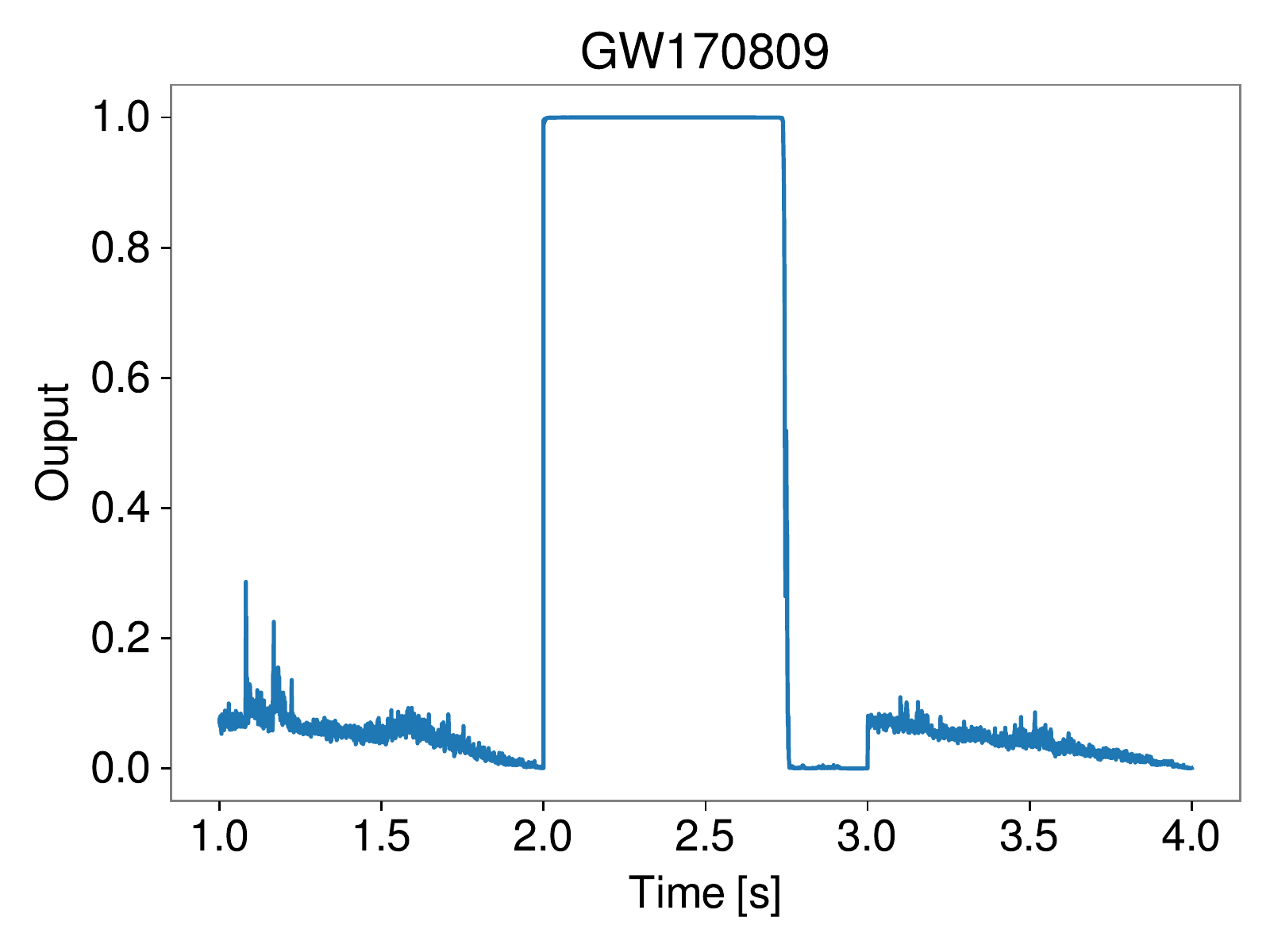}\hfill
}
\caption{Left panel: normalized L--channel spectrograms around three false positives and GW170809. Right panel: as left panel, but now for the detection output of Model \RNum{2}.}
\label{fig:follow_up}
\end{figure*}

We have looked into these three noise triggers to figure out why they were 
singled out by our deep learning ensemble. We present spectrograms and the 
response of Model \RNum{2} to these events in Figure~\ref{fig:follow_up}.
As shown in the left panels of this figure, all three false positives are caused 
by loud glitches in Livingston data. Another interesting result we 
observe in these panels is that the response of our deep learning models to 
these false positives is different to real events, as shown in the bottom 
panels of Figure~\ref{fig:follow_up}; see also Figure~\ref{fig:output}. Note that the response of our neural networks 
for real events is sharp-edged, whereas the neural networks' response 
to noise anomalies is, at best, jagged. This is an additional feature that may be used to 
tell apart real events from other noise anomalies. 

In summary, we have designed a deep learning ensemble that can identify binary 
black hole mergers in O2 and O3 data. Our benchmark analyses indicate that our 
ensemble can process 200 hrs of advanced LIGO noise within 14 hours using 
one node in the HAL cluster, which consists of 4 NVIDIA V100 GPUs. We have found that 
this approach identifies real events in advanced LIGO data, and produces 1 misclassification 
for every 2.7 days of searched data. We have also found that our ensemble can 
generalize to astrophysical signals whose parameters are beyond the parameter space 
used for training, which furnishes evidence for the ability of our models to generalize to 
new signals. 

This new method lays the foundation for the design of a production scale  
deep learning pipeline for gravitational wave searches, which we will present in 
an upcoming publication. 

\section{Conclusions}
\label{sec:end}

We have introduced neural networks that cover a 4-D signal manifold 
that describe quasi-circular, spinning, non-precessing binary black hole mergers. 
We have shown that 
the use of deep learning ensembles enables the detection of O2 and O3 binary 
black hole mergers. We 
have also demonstrated that when this method is applied to hundreds of hours of 
advanced LIGO noise, 
we can identify real events contained in these data nearly ten times faster than 
real-time, with the additional advantage 
of reducing the number 
of false positives to about one for every 2.7 days of searched data. 

Future work will build upon this framework, enlarging the parameter space so as to 
cover a wider range of astrophysical sources that are detectable by advanced 
ground-based detectors, including binary neutron stars and neutron star-black hole mergers, the latter 
being enhanced by early warning detection methods~\cite{wei_warning}.

As data associated with new gravitational wave detections become available through the 
\texttt{Gravitational-Wave Open Science Center}, it will be feasible to better 
tune detection thresholds in these models, which at this point are experimental in 
nature. It may also be possible to start using real events for training purposes, which 
will increase the sensitivity of deep learning searches. In brief, deep learning methods are 
at a tipping point of enabling accelerated gravitational wave detection searches.

\begin{acknowledgments}
We gratefully acknowledge National Science Foundation (NSF) awards OAC-1931561 and OAC-1934757. XH acknowledges support from the National Center for Supercomputing Applications (NCSA) Students Pushing INnovation (SPIN) program. We are grateful to NVIDIA for donating several V100 GPUs that we used for our analysis.

This work utilized resources supported by the NSF's Major Research Instrumentation program, grant OAC-1725729, as well as the University of Illinois at Urbana-Champaign. This work made use of the Illinois Campus Cluster, a computing resource that is operated by the Illinois Campus Cluster Program (ICCP) in conjunction with the NCSA and which is supported by funds from the University of Illinois at Urbana-Champaign. 

This research used resources of the Argonne Leadership Computing Facility, which is a DOE Office of Science User Facility supported under Contract DE-AC02-06CH11357. We also acknowledge NSF grant TG-PHY160053, which provided us access to the Extreme Science and Engineering Discovery Environment (XSEDE) Bridges-AI system. We thank the \href{http://gravity.ncsa.illinois.edu}{NCSA Gravity Group} for useful feedback.
\end{acknowledgments}

%\clearpage 

\begin{widetext}
\appendix
\section{Additional O2 and O3 binary black hole mergers}
\label{sec:apex}

We present the output of our deep learning ensemble for O2 and O3 binary black hole merger 
detections---see Figure~\ref{fig:second_batch}. Note that as we discussed in the main body of the article, 
our approach identifies these events, and produces no false positives over the minutes and hour-long datasets 
containing these real events.

\begin{figure*}[!htb]
\centerline{
\includegraphics[width=0.5\columnwidth]{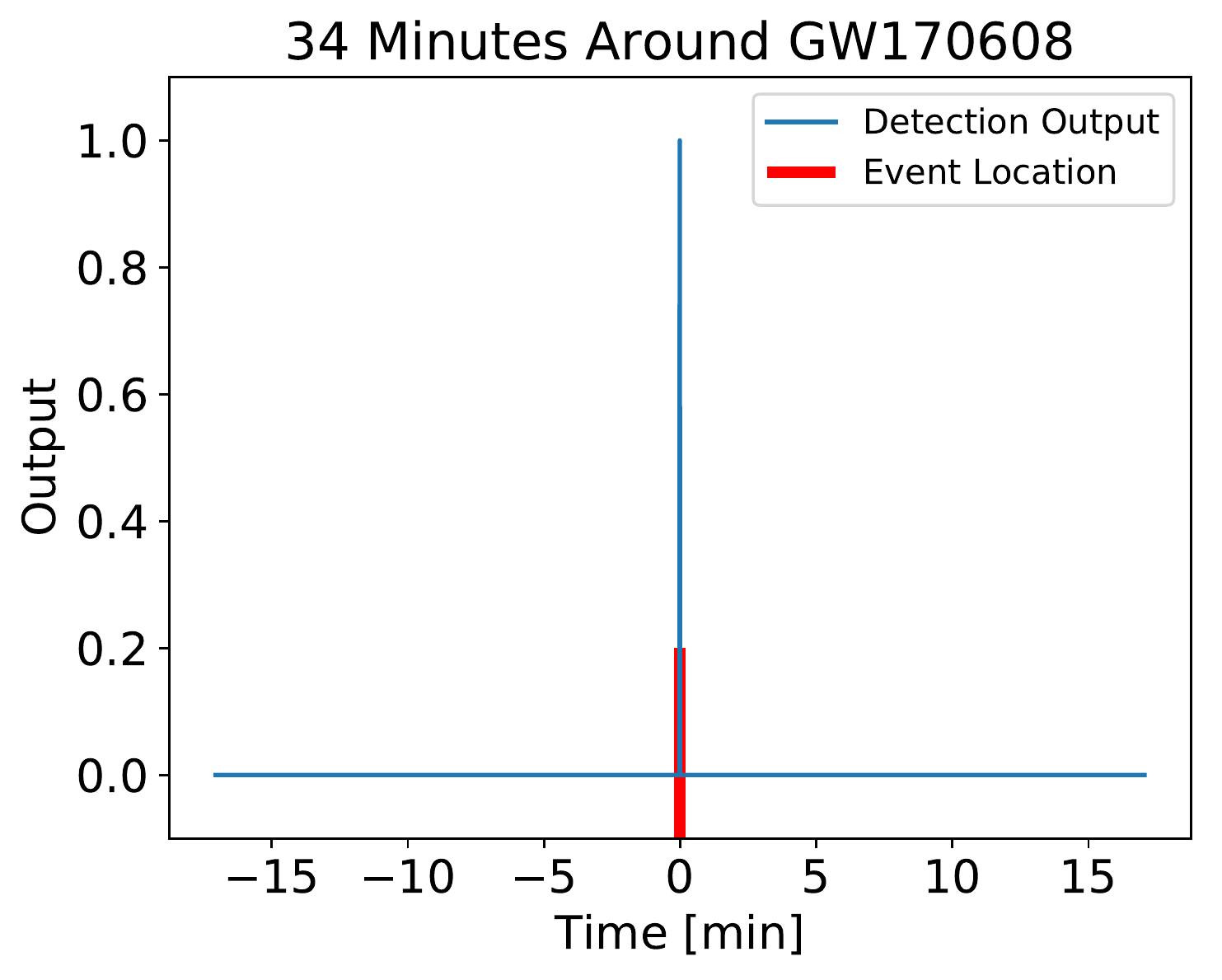}
\includegraphics[width=0.5\columnwidth]{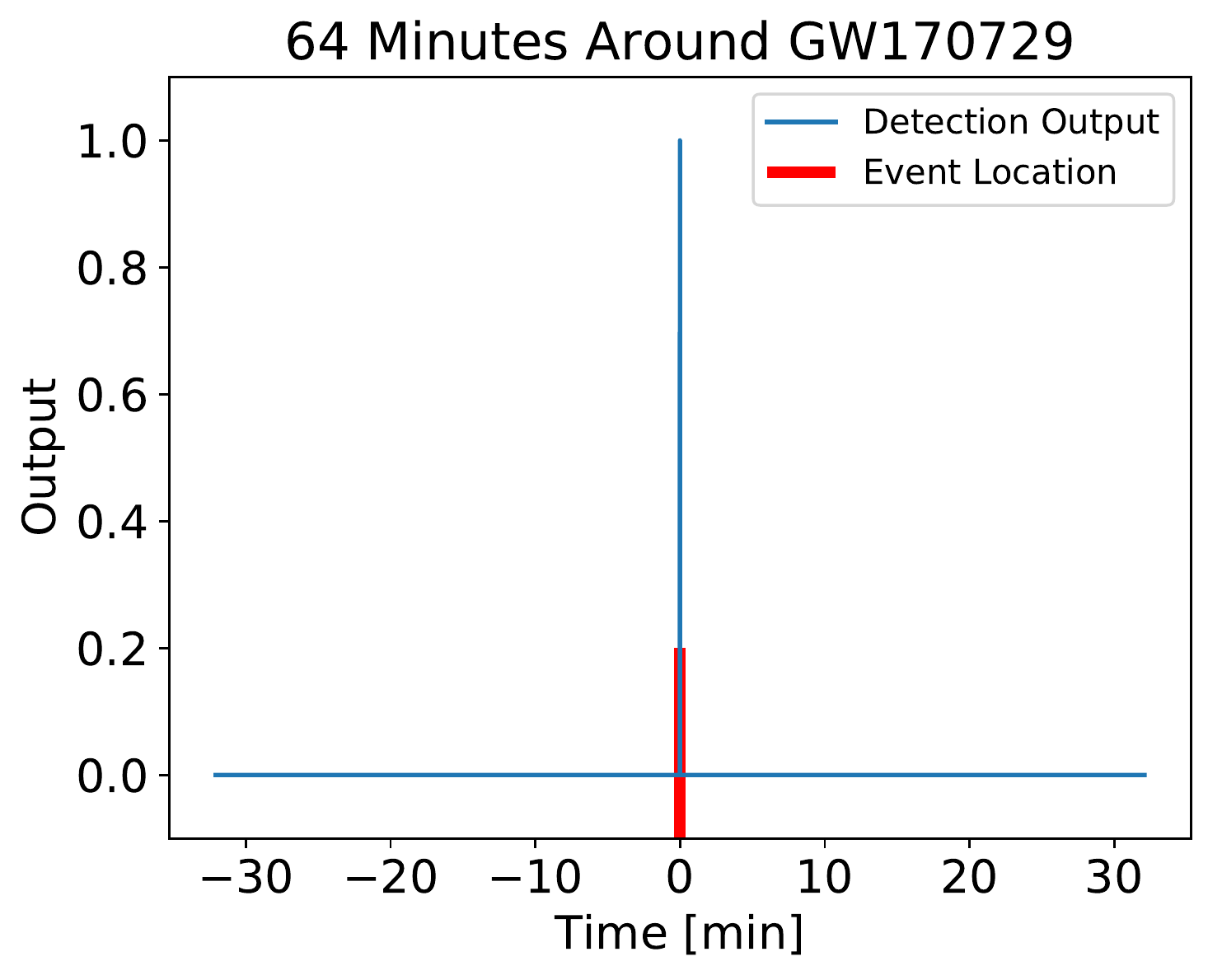}
}
\centerline{
\includegraphics[width=0.5\columnwidth]{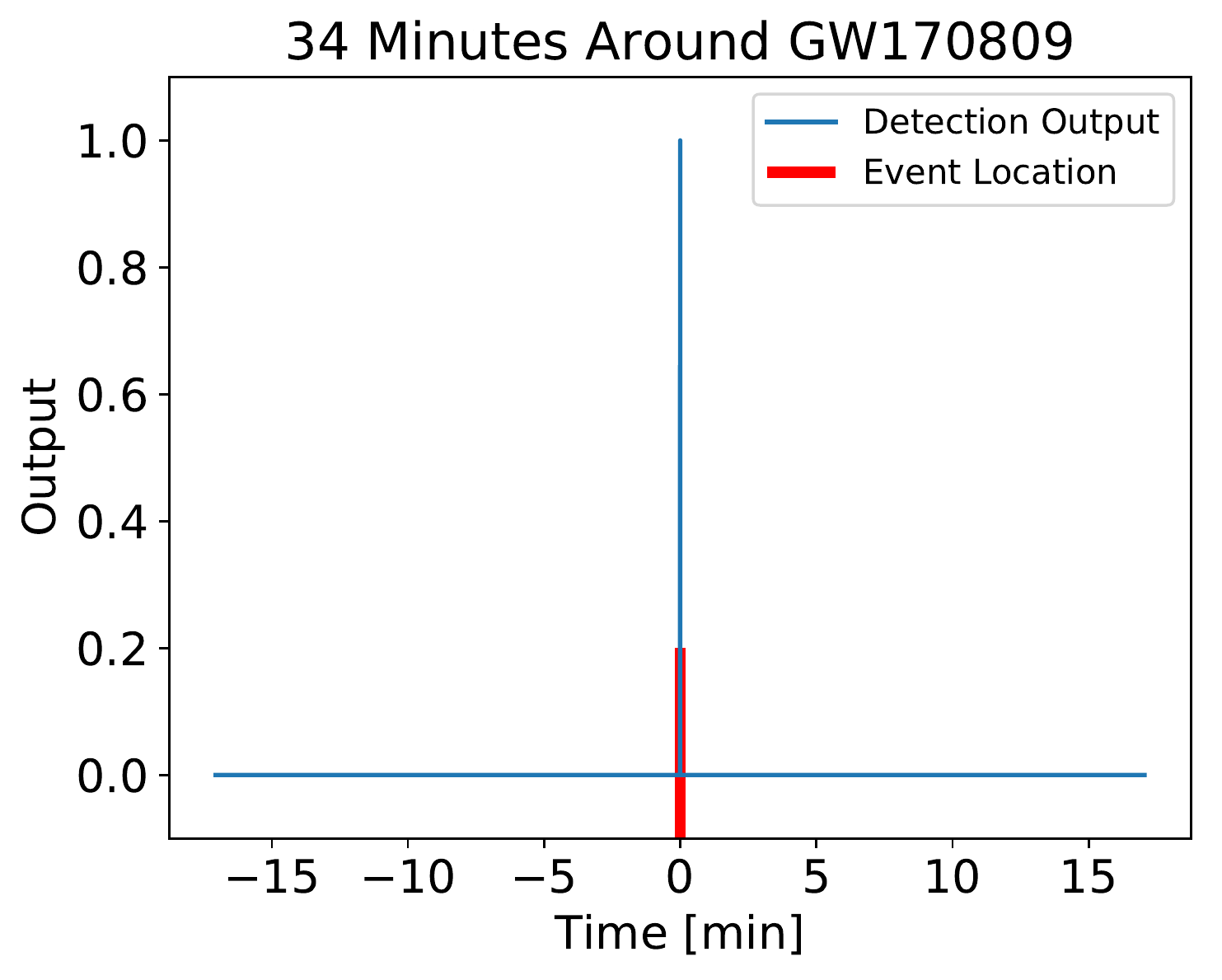}
\includegraphics[width=0.5\columnwidth]{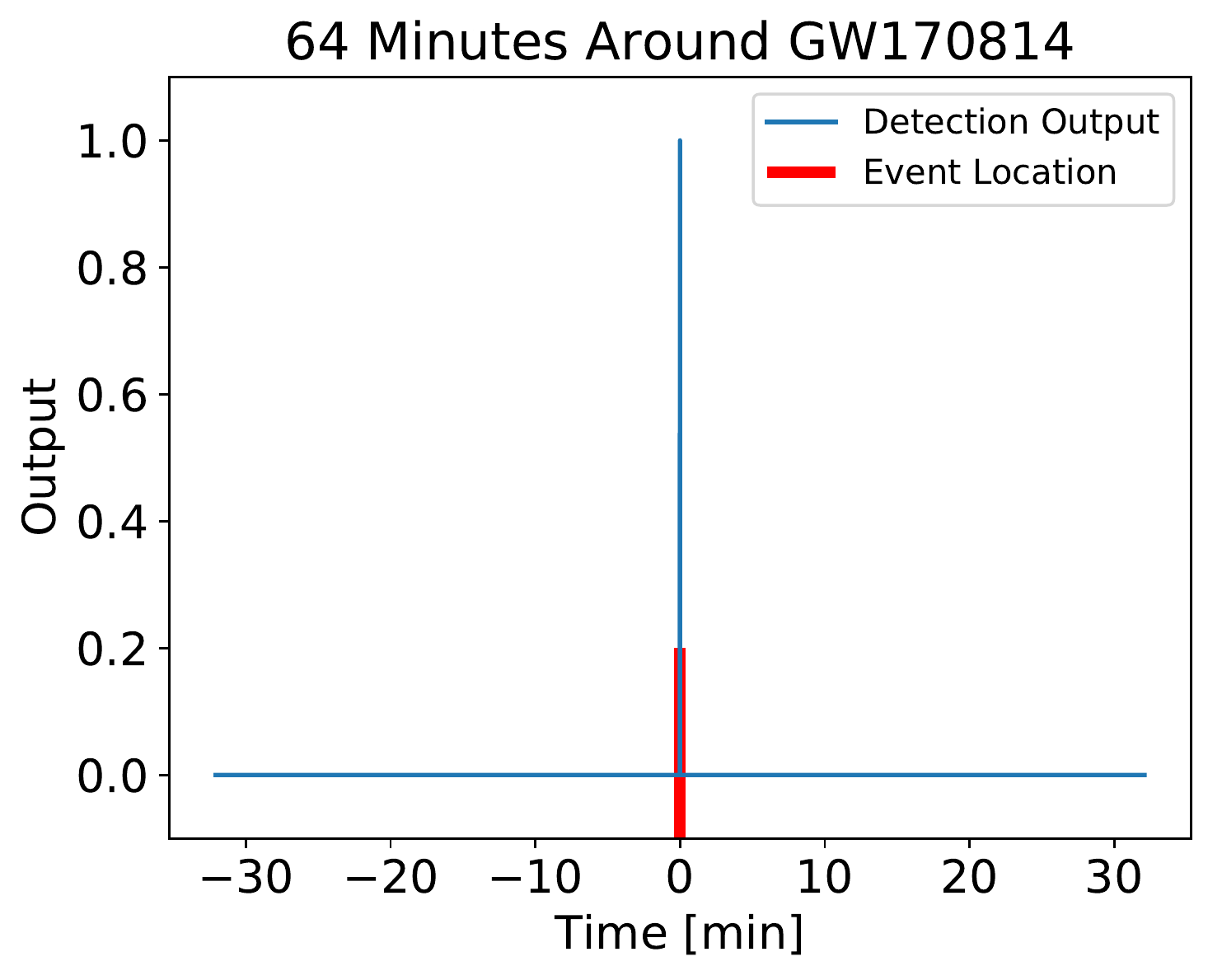}
}
\centerline{
\includegraphics[width=0.5\columnwidth]{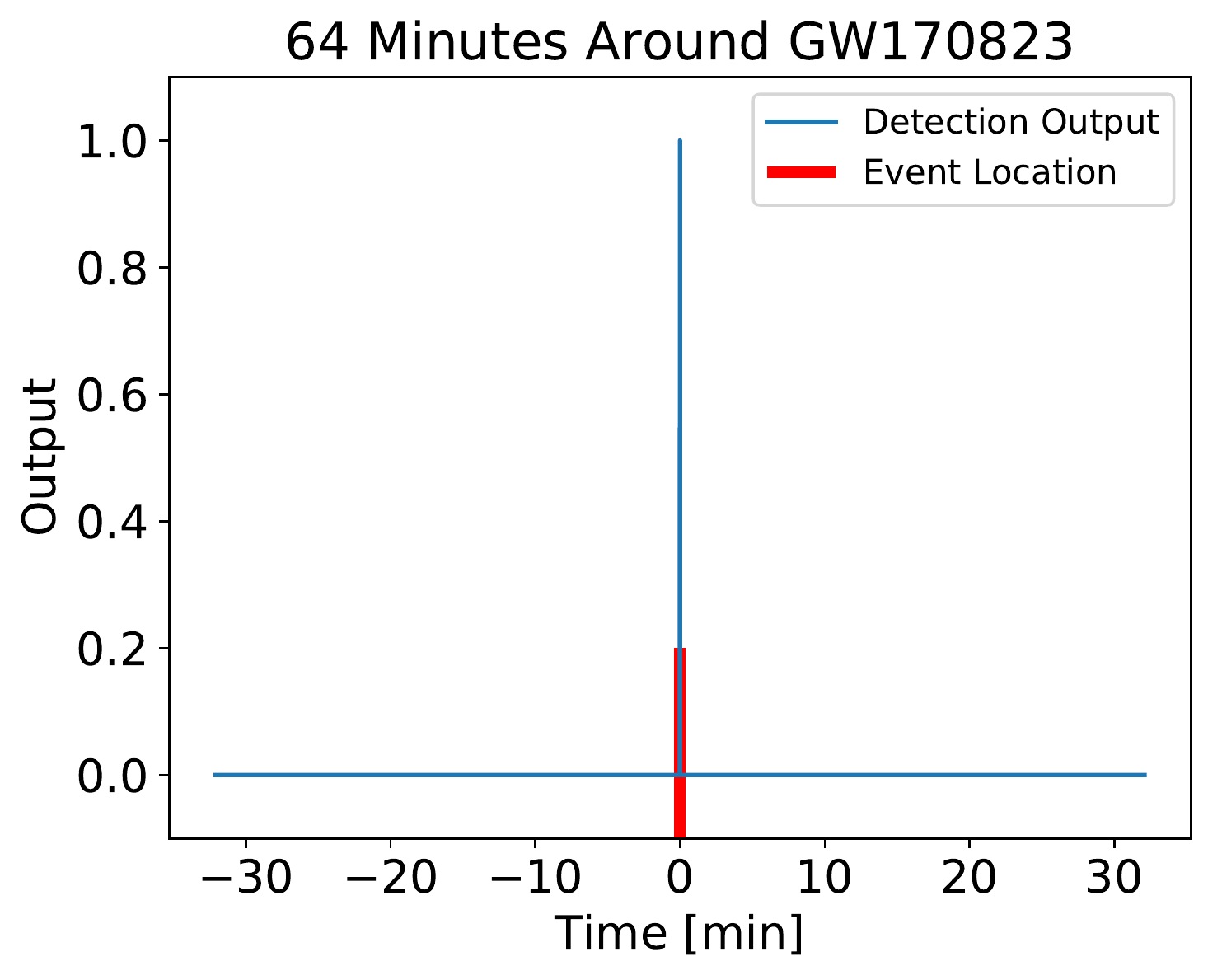}
}
\caption{Detection output of our deep learning ensemble for GW170608, top left; GW170729, top right; 
GW170809, middle left; GW170814, middle right; and GW170823, bottom panel. Notice that our 
ensemble identifies all these 
events with no false positives in minutes and hour-long datasets.}
\label{fig:second_batch}
\end{figure*}
\end{widetext}

%%%%%%%%%%%%%%%%%%%%%%%%%%%%%%%%%%%%%%%%%%%%%

\clearpage

\bibliography{book_references}

\begin{thebibliography}{42}%
\makeatletter
\providecommand \@ifxundefined [1]{%
 \@ifx{#1\undefined}
}%
\providecommand \@ifnum [1]{%
 \ifnum #1\expandafter \@firstoftwo
 \else \expandafter \@secondoftwo
 \fi
}%
\providecommand \@ifx [1]{%
 \ifx #1\expandafter \@firstoftwo
 \else \expandafter \@secondoftwo
 \fi
}%
\providecommand \natexlab [1]{#1}%
\providecommand \enquote  [1]{``#1''}%
\providecommand \bibnamefont  [1]{#1}%
\providecommand \bibfnamefont [1]{#1}%
\providecommand \citenamefont [1]{#1}%
\providecommand \href@noop [0]{\@secondoftwo}%
\providecommand \href [0]{\begingroup \@sanitize@url \@href}%
\providecommand \@href[1]{\@@startlink{#1}\@@href}%
\providecommand \@@href[1]{\endgroup#1\@@endlink}%
\providecommand \@sanitize@url [0]{\catcode `\\12\catcode `\$12\catcode
  `\&12\catcode `\#12\catcode `\^12\catcode `\_12\catcode `\%12\relax}%
\providecommand \@@startlink[1]{}%
\providecommand \@@endlink[0]{}%
\providecommand \url  [0]{\begingroup\@sanitize@url \@url }%
\providecommand \@url [1]{\endgroup\@href {#1}{\urlprefix }}%
\providecommand \urlprefix  [0]{URL }%
\providecommand \Eprint [0]{\href }%
\providecommand \doibase [0]{http://dx.doi.org/}%
\providecommand \selectlanguage [0]{\@gobble}%
\providecommand \bibinfo  [0]{\@secondoftwo}%
\providecommand \bibfield  [0]{\@secondoftwo}%
\providecommand \translation [1]{[#1]}%
\providecommand \BibitemOpen [0]{}%
\providecommand \bibitemStop [0]{}%
\providecommand \bibitemNoStop [0]{.\EOS\space}%
\providecommand \EOS [0]{\spacefactor3000\relax}%
\providecommand \BibitemShut  [1]{\csname bibitem#1\endcsname}%
\let\auto@bib@innerbib\@empty
%</preamble>
\bibitem [{\citenamefont {{The LIGO Scientific Collaboration}}\ \emph
  {et~al.}(2015)\citenamefont {{The LIGO Scientific Collaboration}},
  \citenamefont {{Aasi}} \emph {et~al.}}]{LSC:2015}%
  \BibitemOpen
  \bibfield  {author} {\bibinfo {author} {\bibnamefont {{The LIGO Scientific
  Collaboration}}}, \bibinfo {author} {\bibfnamefont {J.}~\bibnamefont
  {{Aasi}}},  \emph {et~al.},\ }\href {\doibase 10.1088/0264-9381/32/7/074001}
  {\bibfield  {journal} {\bibinfo  {journal} {Classical and Quantum Gravity}\
  }\textbf {\bibinfo {volume} {32}},\ \bibinfo {eid} {074001} (\bibinfo {year}
  {2015})},\ \Eprint {http://arxiv.org/abs/1411.4547} {arXiv:1411.4547 [gr-qc]}
  \BibitemShut {NoStop}%
\bibitem [{\citenamefont {{Acernese}}\ \emph {et~al.}(2015)\citenamefont
  {{Acernese}} \emph {et~al.}}]{Virgo:2015}%
  \BibitemOpen
  \bibfield  {author} {\bibinfo {author} {\bibfnamefont {F.}~\bibnamefont
  {{Acernese}}} \emph {et~al.},\ }\href {\doibase
  10.1088/0264-9381/32/2/024001} {\bibfield  {journal} {\bibinfo  {journal}
  {Classical and Quantum Gravity}\ }\textbf {\bibinfo {volume} {32}},\ \bibinfo
  {eid} {024001} (\bibinfo {year} {2015})}\BibitemShut {NoStop}%
\bibitem [{\citenamefont {Abbott}\ \emph
  {et~al.}(2020{\natexlab{a}})\citenamefont {Abbott} \emph
  {et~al.}}]{Abbott:2020niy}%
  \BibitemOpen
  \bibfield  {author} {\bibinfo {author} {\bibfnamefont {R.}~\bibnamefont
  {Abbott}} \emph {et~al.},\ }\href@noop {} {\  (\bibinfo {year}
  {2020}{\natexlab{a}})},\ \Eprint {http://arxiv.org/abs/2010.14527}
  {arXiv:2010.14527 [gr-qc]} \BibitemShut {NoStop}%
\bibitem [{\citenamefont {George}\ and\ \citenamefont
  {Huerta}(2018{\natexlab{a}})}]{geodf:2017a}%
  \BibitemOpen
  \bibfield  {author} {\bibinfo {author} {\bibfnamefont {D.}~\bibnamefont
  {George}}\ and\ \bibinfo {author} {\bibfnamefont {E.~A.}\ \bibnamefont
  {Huerta}},\ }\href {\doibase 10.1103/PhysRevD.97.044039} {\bibfield
  {journal} {\bibinfo  {journal} {Phys. Rev. D}\ }\textbf {\bibinfo {volume}
  {97}},\ \bibinfo {pages} {044039} (\bibinfo {year} {2018}{\natexlab{a}})},\
  \Eprint {http://arxiv.org/abs/1701.00008} {arXiv:1701.00008 [astro-ph.IM]}
  \BibitemShut {NoStop}%
\bibitem [{\citenamefont {George}\ \emph {et~al.}(2017)\citenamefont {George},
  \citenamefont {Shen},\ and\ \citenamefont {Huerta}}]{George:2017qtr}%
  \BibitemOpen
  \bibfield  {author} {\bibinfo {author} {\bibfnamefont {D.}~\bibnamefont
  {George}}, \bibinfo {author} {\bibfnamefont {H.}~\bibnamefont {Shen}}, \ and\
  \bibinfo {author} {\bibfnamefont {E.}~\bibnamefont {Huerta}},\ }in\
  \href@noop {} {\emph {\bibinfo {booktitle} {{NiPS Summer School 2017}}}}\
  (\bibinfo {year} {2017})\ \Eprint {http://arxiv.org/abs/1711.07468}
  {arXiv:1711.07468 [astro-ph.IM]} \BibitemShut {NoStop}%
\bibitem [{\citenamefont {George}\ and\ \citenamefont
  {Huerta}(2018{\natexlab{b}})}]{GEORGE201864}%
  \BibitemOpen
  \bibfield  {author} {\bibinfo {author} {\bibfnamefont {D.}~\bibnamefont
  {George}}\ and\ \bibinfo {author} {\bibfnamefont {E.}~\bibnamefont
  {Huerta}},\ }\href {\doibase https://doi.org/10.1016/j.physletb.2017.12.053}
  {\bibfield  {journal} {\bibinfo  {journal} {Physics Letters B}\ }\textbf
  {\bibinfo {volume} {778}},\ \bibinfo {pages} {64 } (\bibinfo {year}
  {2018}{\natexlab{b}})}\BibitemShut {NoStop}%
\bibitem [{\citenamefont {{Gabbard}}\ \emph {et~al.}(2018)\citenamefont
  {{Gabbard}}, \citenamefont {{Williams}}, \citenamefont {{Hayes}},\ and\
  \citenamefont {{Messenger}}}]{2018GN}%
  \BibitemOpen
  \bibfield  {author} {\bibinfo {author} {\bibfnamefont {H.}~\bibnamefont
  {{Gabbard}}}, \bibinfo {author} {\bibfnamefont {M.}~\bibnamefont
  {{Williams}}}, \bibinfo {author} {\bibfnamefont {F.}~\bibnamefont {{Hayes}}},
  \ and\ \bibinfo {author} {\bibfnamefont {C.}~\bibnamefont {{Messenger}}},\
  }\href {\doibase 10.1103/PhysRevLett.120.141103} {\bibfield  {journal}
  {\bibinfo  {journal} {Physical Review Letters}\ }\textbf {\bibinfo {volume}
  {120}},\ \bibinfo {eid} {141103} (\bibinfo {year} {2018})},\ \Eprint
  {http://arxiv.org/abs/1712.06041} {arXiv:1712.06041 [astro-ph.IM]}
  \BibitemShut {NoStop}%
\bibitem [{\citenamefont {Skliris}\ \emph {et~al.}(2020)\citenamefont
  {Skliris}, \citenamefont {Norman},\ and\ \citenamefont
  {Sutton}}]{Skliris:2020qax}%
  \BibitemOpen
  \bibfield  {author} {\bibinfo {author} {\bibfnamefont {V.}~\bibnamefont
  {Skliris}}, \bibinfo {author} {\bibfnamefont {M.~R.}\ \bibnamefont {Norman}},
  \ and\ \bibinfo {author} {\bibfnamefont {P.~J.}\ \bibnamefont {Sutton}},\
  }\href@noop {} {\bibfield  {journal} {\bibinfo  {journal} {arXiv preprint
  arXiv:2009.14611}\ } (\bibinfo {year} {2020})}\BibitemShut {NoStop}%
\bibitem [{\citenamefont {Lin}\ and\ \citenamefont {Wu}(2020)}]{Lin:2020aps}%
  \BibitemOpen
  \bibfield  {author} {\bibinfo {author} {\bibfnamefont {Y.-C.}\ \bibnamefont
  {Lin}}\ and\ \bibinfo {author} {\bibfnamefont {J.-H.~P.}\ \bibnamefont
  {Wu}},\ }\href@noop {} {\bibfield  {journal} {\bibinfo  {journal} {arXiv
  preprint arXiv:2007.04176}\ } (\bibinfo {year} {2020})}\BibitemShut {NoStop}%
\bibitem [{\citenamefont {Wang}\ \emph {et~al.}(2020)\citenamefont {Wang},
  \citenamefont {Wu}, \citenamefont {Cao}, \citenamefont {Liu},\ and\
  \citenamefont {Zhu}}]{Wang:2019zaj}%
  \BibitemOpen
  \bibfield  {author} {\bibinfo {author} {\bibfnamefont {H.}~\bibnamefont
  {Wang}}, \bibinfo {author} {\bibfnamefont {S.}~\bibnamefont {Wu}}, \bibinfo
  {author} {\bibfnamefont {Z.}~\bibnamefont {Cao}}, \bibinfo {author}
  {\bibfnamefont {X.}~\bibnamefont {Liu}}, \ and\ \bibinfo {author}
  {\bibfnamefont {J.-Y.}\ \bibnamefont {Zhu}},\ }\href {\doibase
  10.1103/PhysRevD.101.104003} {\bibfield  {journal} {\bibinfo  {journal}
  {Phys. Rev. D}\ }\textbf {\bibinfo {volume} {101}},\ \bibinfo {pages}
  {104003} (\bibinfo {year} {2020})},\ \Eprint
  {http://arxiv.org/abs/1909.13442} {arXiv:1909.13442 [astro-ph.IM]}
  \BibitemShut {NoStop}%
\bibitem [{\citenamefont {Nakano}\ \emph {et~al.}(2019)\citenamefont {Nakano},
  \citenamefont {Narikawa}, \citenamefont {Oohara}, \citenamefont {Sakai},
  \citenamefont {Shinkai}, \citenamefont {Takahashi}, \citenamefont {Tanaka},
  \citenamefont {Uchikata}, \citenamefont {Yamamoto},\ and\ \citenamefont
  {Yamamoto}}]{Nakano:2018vay}%
  \BibitemOpen
  \bibfield  {author} {\bibinfo {author} {\bibfnamefont {H.}~\bibnamefont
  {Nakano}}, \bibinfo {author} {\bibfnamefont {T.}~\bibnamefont {Narikawa}},
  \bibinfo {author} {\bibfnamefont {K.-i.}\ \bibnamefont {Oohara}}, \bibinfo
  {author} {\bibfnamefont {K.}~\bibnamefont {Sakai}}, \bibinfo {author}
  {\bibfnamefont {H.-a.}\ \bibnamefont {Shinkai}}, \bibinfo {author}
  {\bibfnamefont {H.}~\bibnamefont {Takahashi}}, \bibinfo {author}
  {\bibfnamefont {T.}~\bibnamefont {Tanaka}}, \bibinfo {author} {\bibfnamefont
  {N.}~\bibnamefont {Uchikata}}, \bibinfo {author} {\bibfnamefont
  {S.}~\bibnamefont {Yamamoto}}, \ and\ \bibinfo {author} {\bibfnamefont
  {T.~S.}\ \bibnamefont {Yamamoto}},\ }\href {\doibase
  10.1103/PhysRevD.99.124032} {\bibfield  {journal} {\bibinfo  {journal} {Phys.
  Rev. D}\ }\textbf {\bibinfo {volume} {99}},\ \bibinfo {pages} {124032}
  (\bibinfo {year} {2019})},\ \Eprint {http://arxiv.org/abs/1811.06443}
  {arXiv:1811.06443 [gr-qc]} \BibitemShut {NoStop}%
\bibitem [{\citenamefont {Fan}\ \emph {et~al.}(2019)\citenamefont {Fan},
  \citenamefont {Li}, \citenamefont {Li}, \citenamefont {Zhong},\ and\
  \citenamefont {Cao}}]{Fan:2018vgw}%
  \BibitemOpen
  \bibfield  {author} {\bibinfo {author} {\bibfnamefont {X.}~\bibnamefont
  {Fan}}, \bibinfo {author} {\bibfnamefont {J.}~\bibnamefont {Li}}, \bibinfo
  {author} {\bibfnamefont {X.}~\bibnamefont {Li}}, \bibinfo {author}
  {\bibfnamefont {Y.}~\bibnamefont {Zhong}}, \ and\ \bibinfo {author}
  {\bibfnamefont {J.}~\bibnamefont {Cao}},\ }\href {\doibase
  10.1007/s11433-018-9321-7} {\bibfield  {journal} {\bibinfo  {journal} {Sci.
  China Phys. Mech. Astron.}\ }\textbf {\bibinfo {volume} {62}},\ \bibinfo
  {pages} {969512} (\bibinfo {year} {2019})},\ \Eprint
  {http://arxiv.org/abs/1811.01380} {arXiv:1811.01380 [astro-ph.IM]}
  \BibitemShut {NoStop}%
\bibitem [{\citenamefont {Li}\ \emph {et~al.}(2020)\citenamefont {Li},
  \citenamefont {Babu}, \citenamefont {Yu},\ and\ \citenamefont
  {Fan}}]{Li:2017chi}%
  \BibitemOpen
  \bibfield  {author} {\bibinfo {author} {\bibfnamefont {X.-R.}\ \bibnamefont
  {Li}}, \bibinfo {author} {\bibfnamefont {G.}~\bibnamefont {Babu}}, \bibinfo
  {author} {\bibfnamefont {W.-L.}\ \bibnamefont {Yu}}, \ and\ \bibinfo {author}
  {\bibfnamefont {X.-L.}\ \bibnamefont {Fan}},\ }\href {\doibase
  10.1007/s11467-020-0966-4} {\bibfield  {journal} {\bibinfo  {journal} {Front.
  Phys. (Beijing)}\ }\textbf {\bibinfo {volume} {15}},\ \bibinfo {pages}
  {54501} (\bibinfo {year} {2020})},\ \Eprint {http://arxiv.org/abs/1712.00356}
  {arXiv:1712.00356 [astro-ph.IM]} \BibitemShut {NoStop}%
\bibitem [{\citenamefont {Deighan}\ \emph {et~al.}(2020)\citenamefont
  {Deighan}, \citenamefont {Field}, \citenamefont {Capano},\ and\ \citenamefont
  {Khanna}}]{Deighan:2020gtp}%
  \BibitemOpen
  \bibfield  {author} {\bibinfo {author} {\bibfnamefont {D.~S.}\ \bibnamefont
  {Deighan}}, \bibinfo {author} {\bibfnamefont {S.~E.}\ \bibnamefont {Field}},
  \bibinfo {author} {\bibfnamefont {C.~D.}\ \bibnamefont {Capano}}, \ and\
  \bibinfo {author} {\bibfnamefont {G.}~\bibnamefont {Khanna}},\ }\href@noop {}
  {\  (\bibinfo {year} {2020})},\ \Eprint {http://arxiv.org/abs/2010.04340}
  {arXiv:2010.04340 [gr-qc]} \BibitemShut {NoStop}%
\bibitem [{\citenamefont {Miller}\ \emph {et~al.}(2019)\citenamefont {Miller}
  \emph {et~al.}}]{Miller:2019jtp}%
  \BibitemOpen
  \bibfield  {author} {\bibinfo {author} {\bibfnamefont {A.~L.}\ \bibnamefont
  {Miller}} \emph {et~al.},\ }\href {\doibase 10.1103/PhysRevD.100.062005}
  {\bibfield  {journal} {\bibinfo  {journal} {Phys. Rev. D}\ }\textbf {\bibinfo
  {volume} {100}},\ \bibinfo {pages} {062005} (\bibinfo {year} {2019})},\
  \Eprint {http://arxiv.org/abs/1909.02262} {arXiv:1909.02262 [astro-ph.IM]}
  \BibitemShut {NoStop}%
\bibitem [{\citenamefont {Krastev}(2020)}]{Krastev:2019koe}%
  \BibitemOpen
  \bibfield  {author} {\bibinfo {author} {\bibfnamefont {P.~G.}\ \bibnamefont
  {Krastev}},\ }\href {\doibase 10.1016/j.physletb.2020.135330} {\bibfield
  {journal} {\bibinfo  {journal} {Phys. Lett. B}\ }\textbf {\bibinfo {volume}
  {803}},\ \bibinfo {pages} {135330} (\bibinfo {year} {2020})},\ \Eprint
  {http://arxiv.org/abs/1908.03151} {arXiv:1908.03151 [astro-ph.IM]}
  \BibitemShut {NoStop}%
\bibitem [{\citenamefont {{Sch{\"a}fer}}\ \emph {et~al.}(2020)\citenamefont
  {{Sch{\"a}fer}}, \citenamefont {{Ohme}},\ and\ \citenamefont
  {{Nitz}}}]{2020PhRvD.102f3015S}%
  \BibitemOpen
  \bibfield  {author} {\bibinfo {author} {\bibfnamefont {M.~B.}\ \bibnamefont
  {{Sch{\"a}fer}}}, \bibinfo {author} {\bibfnamefont {F.}~\bibnamefont
  {{Ohme}}}, \ and\ \bibinfo {author} {\bibfnamefont {A.~H.}\ \bibnamefont
  {{Nitz}}},\ }\href {\doibase 10.1103/PhysRevD.102.063015} {\bibfield
  {journal} {\bibinfo  {journal} {\prd}\ }\textbf {\bibinfo {volume} {102}},\
  \bibinfo {eid} {063015} (\bibinfo {year} {2020})},\ \Eprint
  {http://arxiv.org/abs/2006.01509} {arXiv:2006.01509 [astro-ph.HE]}
  \BibitemShut {NoStop}%
\bibitem [{\citenamefont {Dreissigacker}\ and\ \citenamefont
  {Prix}(2020)}]{Dreissigacker:2020xfr}%
  \BibitemOpen
  \bibfield  {author} {\bibinfo {author} {\bibfnamefont {C.}~\bibnamefont
  {Dreissigacker}}\ and\ \bibinfo {author} {\bibfnamefont {R.}~\bibnamefont
  {Prix}},\ }\href {\doibase 10.1103/PhysRevD.102.022005} {\bibfield  {journal}
  {\bibinfo  {journal} {Phys. Rev. D}\ }\textbf {\bibinfo {volume} {102}},\
  \bibinfo {pages} {022005} (\bibinfo {year} {2020})},\ \Eprint
  {http://arxiv.org/abs/2005.04140} {arXiv:2005.04140 [gr-qc]} \BibitemShut
  {NoStop}%
\bibitem [{\citenamefont {Khan}\ \emph {et~al.}(2020)\citenamefont {Khan},
  \citenamefont {Huerta},\ and\ \citenamefont {Das}}]{Khan:2020foe}%
  \BibitemOpen
  \bibfield  {author} {\bibinfo {author} {\bibfnamefont {A.}~\bibnamefont
  {Khan}}, \bibinfo {author} {\bibfnamefont {E.}~\bibnamefont {Huerta}}, \ and\
  \bibinfo {author} {\bibfnamefont {A.}~\bibnamefont {Das}},\ }\href {\doibase
  10.1016/j.physletb.2020.135628} {\bibfield  {journal} {\bibinfo  {journal}
  {Phys. Lett. B}\ }\textbf {\bibinfo {volume} {808}},\ \bibinfo {pages} {0370}
  (\bibinfo {year} {2020})},\ \Eprint {http://arxiv.org/abs/2004.09524}
  {arXiv:2004.09524 [gr-qc]} \BibitemShut {NoStop}%
\bibitem [{\citenamefont {Dreissigacker}\ \emph {et~al.}(2019)\citenamefont
  {Dreissigacker}, \citenamefont {Sharma}, \citenamefont {Messenger},
  \citenamefont {Zhao},\ and\ \citenamefont {Prix}}]{Dreissigacker:2019edy}%
  \BibitemOpen
  \bibfield  {author} {\bibinfo {author} {\bibfnamefont {C.}~\bibnamefont
  {Dreissigacker}}, \bibinfo {author} {\bibfnamefont {R.}~\bibnamefont
  {Sharma}}, \bibinfo {author} {\bibfnamefont {C.}~\bibnamefont {Messenger}},
  \bibinfo {author} {\bibfnamefont {R.}~\bibnamefont {Zhao}}, \ and\ \bibinfo
  {author} {\bibfnamefont {R.}~\bibnamefont {Prix}},\ }\href {\doibase
  10.1103/PhysRevD.100.044009} {\bibfield  {journal} {\bibinfo  {journal}
  {Phys. Rev. D}\ }\textbf {\bibinfo {volume} {100}},\ \bibinfo {pages}
  {044009} (\bibinfo {year} {2019})},\ \Eprint
  {http://arxiv.org/abs/1904.13291} {arXiv:1904.13291 [gr-qc]} \BibitemShut
  {NoStop}%
\bibitem [{\citenamefont {{Beheshtipour}}\ and\ \citenamefont
  {{Papa}}(2020)}]{2020PhRvD.101f4009B}%
  \BibitemOpen
  \bibfield  {author} {\bibinfo {author} {\bibfnamefont {B.}~\bibnamefont
  {{Beheshtipour}}}\ and\ \bibinfo {author} {\bibfnamefont {M.~A.}\
  \bibnamefont {{Papa}}},\ }\href {\doibase 10.1103/PhysRevD.101.064009}
  {\bibfield  {journal} {\bibinfo  {journal} {\prd}\ }\textbf {\bibinfo
  {volume} {101}},\ \bibinfo {eid} {064009} (\bibinfo {year} {2020})},\ \Eprint
  {http://arxiv.org/abs/2001.03116} {arXiv:2001.03116 [gr-qc]} \BibitemShut
  {NoStop}%
\bibitem [{\citenamefont {{Skliris}}\ \emph {et~al.}(2020)\citenamefont
  {{Skliris}}, \citenamefont {{Norman}},\ and\ \citenamefont
  {{Sutton}}}]{2020arXiv200914611S}%
  \BibitemOpen
  \bibfield  {author} {\bibinfo {author} {\bibfnamefont {V.}~\bibnamefont
  {{Skliris}}}, \bibinfo {author} {\bibfnamefont {M.~R.~K.}\ \bibnamefont
  {{Norman}}}, \ and\ \bibinfo {author} {\bibfnamefont {P.~J.}\ \bibnamefont
  {{Sutton}}},\ }\href@noop {} {\bibfield  {journal} {\bibinfo  {journal}
  {arXiv e-prints}\ ,\ \bibinfo {eid} {arXiv:2009.14611}} (\bibinfo {year}
  {2020})},\ \Eprint {http://arxiv.org/abs/2009.14611} {arXiv:2009.14611
  [astro-ph.IM]} \BibitemShut {NoStop}%
\bibitem [{\citenamefont {Khan}\ and\ \citenamefont
  {Green}(2020)}]{Khan:2020fso}%
  \BibitemOpen
  \bibfield  {author} {\bibinfo {author} {\bibfnamefont {S.}~\bibnamefont
  {Khan}}\ and\ \bibinfo {author} {\bibfnamefont {R.}~\bibnamefont {Green}},\
  }\href@noop {} {\bibfield  {journal} {\bibinfo  {journal} {arXiv preprint
  arXiv:2008.12932}\ } (\bibinfo {year} {2020})}\BibitemShut {NoStop}%
\bibitem [{\citenamefont {Chua}\ \emph {et~al.}(2019)\citenamefont {Chua},
  \citenamefont {Galley},\ and\ \citenamefont
  {Vallisneri}}]{PhysRevLett.122.211101}%
  \BibitemOpen
  \bibfield  {author} {\bibinfo {author} {\bibfnamefont {A.~J.~K.}\
  \bibnamefont {Chua}}, \bibinfo {author} {\bibfnamefont {C.~R.}\ \bibnamefont
  {Galley}}, \ and\ \bibinfo {author} {\bibfnamefont {M.}~\bibnamefont
  {Vallisneri}},\ }\href {\doibase 10.1103/PhysRevLett.122.211101} {\bibfield
  {journal} {\bibinfo  {journal} {Phys. Rev. Lett.}\ }\textbf {\bibinfo
  {volume} {122}},\ \bibinfo {pages} {211101} (\bibinfo {year}
  {2019})}\BibitemShut {NoStop}%
\bibitem [{\citenamefont {{Wei}}\ and\ \citenamefont
  {{Huerta}}(2020)}]{wei_warning}%
  \BibitemOpen
  \bibfield  {author} {\bibinfo {author} {\bibfnamefont {W.}~\bibnamefont
  {{Wei}}}\ and\ \bibinfo {author} {\bibfnamefont {E.~A.}\ \bibnamefont
  {{Huerta}}},\ }\href@noop {} {\bibfield  {journal} {\bibinfo  {journal}
  {arXiv e-prints}\ ,\ \bibinfo {eid} {arXiv:2010.09751}} (\bibinfo {year}
  {2020})},\ \Eprint {http://arxiv.org/abs/2010.09751} {arXiv:2010.09751
  [gr-qc]} \BibitemShut {NoStop}%
\bibitem [{\citenamefont {Vallisneri}\ \emph {et~al.}(2015)\citenamefont
  {Vallisneri}, \citenamefont {Kanner}, \citenamefont {Williams}, \citenamefont
  {Weinstein},\ and\ \citenamefont {Stephens}}]{Vallisneri:2014vxa}%
  \BibitemOpen
  \bibfield  {author} {\bibinfo {author} {\bibfnamefont {M.}~\bibnamefont
  {Vallisneri}}, \bibinfo {author} {\bibfnamefont {J.}~\bibnamefont {Kanner}},
  \bibinfo {author} {\bibfnamefont {R.}~\bibnamefont {Williams}}, \bibinfo
  {author} {\bibfnamefont {A.}~\bibnamefont {Weinstein}}, \ and\ \bibinfo
  {author} {\bibfnamefont {B.}~\bibnamefont {Stephens}},\ }\bibfield
  {booktitle} {\emph {\bibinfo {booktitle} {{Proceedings, 10th International
  LISA Symposium: Gainesville, Florida, USA, May 18-23, 2014}}},\ }\href
  {\doibase 10.1088/1742-6596/610/1/012021} {\bibfield  {journal} {\bibinfo
  {journal} {J. Phys. Conf. Ser.}\ }\textbf {\bibinfo {volume} {610}},\
  \bibinfo {pages} {012021} (\bibinfo {year} {2015})},\ \Eprint
  {http://arxiv.org/abs/1410.4839} {arXiv:1410.4839 [gr-qc]} \BibitemShut
  {NoStop}%
%%CITATION = ARXIV:1410.4839;%%
\bibitem [{\citenamefont {{Raissi}}\ \emph
  {et~al.}(2017{\natexlab{a}})\citenamefont {{Raissi}}, \citenamefont
  {{Perdikaris}},\ and\ \citenamefont {{Karniadakis}}}]{pidl1}%
  \BibitemOpen
  \bibfield  {author} {\bibinfo {author} {\bibfnamefont {M.}~\bibnamefont
  {{Raissi}}}, \bibinfo {author} {\bibfnamefont {P.}~\bibnamefont
  {{Perdikaris}}}, \ and\ \bibinfo {author} {\bibfnamefont {G.~E.}\
  \bibnamefont {{Karniadakis}}},\ }\href@noop {} {\bibfield  {journal}
  {\bibinfo  {journal} {arXiv e-prints}\ ,\ \bibinfo {eid} {arXiv:1711.10561}}
  (\bibinfo {year} {2017}{\natexlab{a}})},\ \Eprint
  {http://arxiv.org/abs/1711.10561} {arXiv:1711.10561 [cs.AI]} \BibitemShut
  {NoStop}%
\bibitem [{\citenamefont {{Raissi}}\ \emph
  {et~al.}(2017{\natexlab{b}})\citenamefont {{Raissi}}, \citenamefont
  {{Perdikaris}},\ and\ \citenamefont {{Karniadakis}}}]{pidl2}%
  \BibitemOpen
  \bibfield  {author} {\bibinfo {author} {\bibfnamefont {M.}~\bibnamefont
  {{Raissi}}}, \bibinfo {author} {\bibfnamefont {P.}~\bibnamefont
  {{Perdikaris}}}, \ and\ \bibinfo {author} {\bibfnamefont {G.~E.}\
  \bibnamefont {{Karniadakis}}},\ }\href@noop {} {\bibfield  {journal}
  {\bibinfo  {journal} {arXiv e-prints}\ ,\ \bibinfo {eid} {arXiv:1711.10566}}
  (\bibinfo {year} {2017}{\natexlab{b}})},\ \Eprint
  {http://arxiv.org/abs/1711.10566} {arXiv:1711.10566 [cs.AI]} \BibitemShut
  {NoStop}%
\bibitem [{\citenamefont {Wei}\ and\ \citenamefont
  {Huerta}(2020)}]{Wei:2019zlc}%
  \BibitemOpen
  \bibfield  {author} {\bibinfo {author} {\bibfnamefont {W.}~\bibnamefont
  {Wei}}\ and\ \bibinfo {author} {\bibfnamefont {E.~A.}\ \bibnamefont
  {Huerta}},\ }\href {\doibase 10.1016/j.physletb.2019.135081} {\bibfield
  {journal} {\bibinfo  {journal} {Phys. Lett.}\ }\textbf {\bibinfo {volume}
  {B800}},\ \bibinfo {pages} {135081} (\bibinfo {year} {2020})},\ \Eprint
  {http://arxiv.org/abs/1901.00869} {arXiv:1901.00869 [gr-qc]} \BibitemShut
  {NoStop}%
%%CITATION = ARXIV:1901.00869;%%
\bibitem [{\citenamefont {{Khan Asad, E. A. Huerta and Arnav
  Das}}(2020)}]{dlhubmodel1}%
  \BibitemOpen
  \bibfield  {author} {\bibinfo {author} {\bibnamefont {{Khan Asad, E. A.
  Huerta and Arnav Das}}},\ }\href@noop {} {\enquote {\bibinfo {title} {{A deep
  learning model to characterize the signal manifold of quasi-circular,
  spinning, non-precessing binary black hole mergers}},}\ } (\bibinfo {year}
  {2020}),\ \bibinfo {note}
  {\url{https://doi.org/10.26311/8wnt-3343}}\BibitemShut {NoStop}%
\bibitem [{\citenamefont {{van den Oord}}\ \emph {et~al.}(2016)\citenamefont
  {{van den Oord}}, \citenamefont {{Dieleman}}, \citenamefont {{Zen}},
  \citenamefont {{Simonyan}}, \citenamefont {{Vinyals}}, \citenamefont
  {{Graves}}, \citenamefont {{Kalchbrenner}}, \citenamefont {{Senior}},\ and\
  \citenamefont {{Kavukcuoglu}}}]{2016wavenet}%
  \BibitemOpen
  \bibfield  {author} {\bibinfo {author} {\bibfnamefont {A.}~\bibnamefont {{van
  den Oord}}}, \bibinfo {author} {\bibfnamefont {S.}~\bibnamefont
  {{Dieleman}}}, \bibinfo {author} {\bibfnamefont {H.}~\bibnamefont {{Zen}}},
  \bibinfo {author} {\bibfnamefont {K.}~\bibnamefont {{Simonyan}}}, \bibinfo
  {author} {\bibfnamefont {O.}~\bibnamefont {{Vinyals}}}, \bibinfo {author}
  {\bibfnamefont {A.}~\bibnamefont {{Graves}}}, \bibinfo {author}
  {\bibfnamefont {N.}~\bibnamefont {{Kalchbrenner}}}, \bibinfo {author}
  {\bibfnamefont {A.}~\bibnamefont {{Senior}}}, \ and\ \bibinfo {author}
  {\bibfnamefont {K.}~\bibnamefont {{Kavukcuoglu}}},\ }\href@noop {} {\bibfield
   {journal} {\bibinfo  {journal} {arXiv e-prints}\ ,\ \bibinfo {eid}
  {arXiv:1609.03499}} (\bibinfo {year} {2016})},\ \Eprint
  {http://arxiv.org/abs/1609.03499} {arXiv:1609.03499 [cs.SD]} \BibitemShut
  {NoStop}%
\bibitem [{\citenamefont {Krizhevsky}\ \emph {et~al.}(2012)\citenamefont
  {Krizhevsky}, \citenamefont {Sutskever},\ and\ \citenamefont
  {Hinton}}]{krizhevsky2012imagenet}%
  \BibitemOpen
  \bibfield  {author} {\bibinfo {author} {\bibfnamefont {A.}~\bibnamefont
  {Krizhevsky}}, \bibinfo {author} {\bibfnamefont {I.}~\bibnamefont
  {Sutskever}}, \ and\ \bibinfo {author} {\bibfnamefont {G.~E.}\ \bibnamefont
  {Hinton}},\ }in\ \href
  {http://papers.nips.cc/paper/4824-imagenet-classification-with-deep-convolutional-neural-networks}
  {\emph {\bibinfo {booktitle} {Advances in neural information processing
  systems}}}\ (\bibinfo {year} {2012})\ pp.\ \bibinfo {pages}
  {1097--1105}\BibitemShut {NoStop}%
\bibitem [{\citenamefont {van~den Oord}\ \emph {et~al.}(2016)\citenamefont
  {van~den Oord}, \citenamefont {Dieleman}, \citenamefont {Zen}, \citenamefont
  {Simonyan}, \citenamefont {Vinyals}, \citenamefont {Graves}, \citenamefont
  {Kalchbrenner}, \citenamefont {Senior},\ and\ \citenamefont
  {Kavukcuoglu}}]{oord2016wavenet}%
  \BibitemOpen
  \bibfield  {author} {\bibinfo {author} {\bibfnamefont {A.}~\bibnamefont
  {van~den Oord}}, \bibinfo {author} {\bibfnamefont {S.}~\bibnamefont
  {Dieleman}}, \bibinfo {author} {\bibfnamefont {H.}~\bibnamefont {Zen}},
  \bibinfo {author} {\bibfnamefont {K.}~\bibnamefont {Simonyan}}, \bibinfo
  {author} {\bibfnamefont {O.}~\bibnamefont {Vinyals}}, \bibinfo {author}
  {\bibfnamefont {A.}~\bibnamefont {Graves}}, \bibinfo {author} {\bibfnamefont
  {N.}~\bibnamefont {Kalchbrenner}}, \bibinfo {author} {\bibfnamefont
  {A.}~\bibnamefont {Senior}}, \ and\ \bibinfo {author} {\bibfnamefont
  {K.}~\bibnamefont {Kavukcuoglu}},\ }\href@noop {} {\enquote {\bibinfo {title}
  {Wavenet: A generative model for raw audio},}\ } (\bibinfo {year} {2016}),\
  \Eprint {http://arxiv.org/abs/1609.03499} {arXiv:1609.03499 [cs.SD]}
  \BibitemShut {NoStop}%
\bibitem [{\citenamefont {He}\ \emph {et~al.}(2016)\citenamefont {He},
  \citenamefont {Zhang}, \citenamefont {Ren},\ and\ \citenamefont
  {Sun}}]{he2016deep}%
  \BibitemOpen
  \bibfield  {author} {\bibinfo {author} {\bibfnamefont {K.}~\bibnamefont
  {He}}, \bibinfo {author} {\bibfnamefont {X.}~\bibnamefont {Zhang}}, \bibinfo
  {author} {\bibfnamefont {S.}~\bibnamefont {Ren}}, \ and\ \bibinfo {author}
  {\bibfnamefont {J.}~\bibnamefont {Sun}},\ }in\ \href@noop {} {\emph {\bibinfo
  {booktitle} {Proceedings of the IEEE conference on computer vision and
  pattern recognition}}}\ (\bibinfo {year} {2016})\ pp.\ \bibinfo {pages}
  {770--778}\BibitemShut {NoStop}%
\bibitem [{\citenamefont {{Pan}}\ \emph {et~al.}(2014)\citenamefont {{Pan}},
  \citenamefont {{Buonanno}}, \citenamefont {{Taracchini}}, \citenamefont
  {{Kidder}}, \citenamefont {{Mrou{\'e}}}, \citenamefont {{Pfeiffer}},
  \citenamefont {{Scheel}},\ and\ \citenamefont {{Szil{\'a}gyi}}}]{seobnrv3}%
  \BibitemOpen
  \bibfield  {author} {\bibinfo {author} {\bibfnamefont {Y.}~\bibnamefont
  {{Pan}}}, \bibinfo {author} {\bibfnamefont {A.}~\bibnamefont {{Buonanno}}},
  \bibinfo {author} {\bibfnamefont {A.}~\bibnamefont {{Taracchini}}}, \bibinfo
  {author} {\bibfnamefont {L.~E.}\ \bibnamefont {{Kidder}}}, \bibinfo {author}
  {\bibfnamefont {A.~H.}\ \bibnamefont {{Mrou{\'e}}}}, \bibinfo {author}
  {\bibfnamefont {H.~P.}\ \bibnamefont {{Pfeiffer}}}, \bibinfo {author}
  {\bibfnamefont {M.~A.}\ \bibnamefont {{Scheel}}}, \ and\ \bibinfo {author}
  {\bibfnamefont {B.}~\bibnamefont {{Szil{\'a}gyi}}},\ }\href {\doibase
  10.1103/PhysRevD.89.084006} {\bibfield  {journal} {\bibinfo  {journal}
  {\prd}\ }\textbf {\bibinfo {volume} {89}},\ \bibinfo {eid} {084006} (\bibinfo
  {year} {2014})},\ \Eprint {http://arxiv.org/abs/1307.6232} {arXiv:1307.6232
  [gr-qc]} \BibitemShut {NoStop}%
\bibitem [{\citenamefont {Abbott}\ \emph
  {et~al.}(2020{\natexlab{b}})\citenamefont {Abbott} \emph
  {et~al.}}]{Abbott:2020tfl}%
  \BibitemOpen
  \bibfield  {author} {\bibinfo {author} {\bibfnamefont {R.}~\bibnamefont
  {Abbott}} \emph {et~al.} (\bibinfo {collaboration} {LIGO Scientific,
  Virgo}),\ }\href {\doibase 10.1103/PhysRevLett.125.101102} {\bibfield
  {journal} {\bibinfo  {journal} {Phys. Rev. Lett.}\ }\textbf {\bibinfo
  {volume} {125}},\ \bibinfo {pages} {101102} (\bibinfo {year}
  {2020}{\natexlab{b}})},\ \Eprint {http://arxiv.org/abs/2009.01075}
  {arXiv:2009.01075 [gr-qc]} \BibitemShut {NoStop}%
\bibitem [{\citenamefont {{Usman}}\ \emph {et~al.}(2016)\citenamefont {{Usman}}
  \emph {et~al.}}]{2016CQGra..33u5004U}%
  \BibitemOpen
  \bibfield  {author} {\bibinfo {author} {\bibfnamefont {S.~A.}\ \bibnamefont
  {{Usman}}} \emph {et~al.},\ }\href {\doibase 10.1088/0264-9381/33/21/215004}
  {\bibfield  {journal} {\bibinfo  {journal} {Classical and Quantum Gravity}\
  }\textbf {\bibinfo {volume} {33}},\ \bibinfo {eid} {215004} (\bibinfo {year}
  {2016})},\ \Eprint {http://arxiv.org/abs/1508.02357} {arXiv:1508.02357
  [gr-qc]} \BibitemShut {NoStop}%
\bibitem [{\citenamefont {{XSEDE}}()}]{bridgesai}%
  \BibitemOpen
  \bibfield  {author} {\bibinfo {author} {\bibnamefont {{XSEDE}}},\ }\href@noop
  {} {\enquote {\bibinfo {title} {{Bridges-AI}},}\ }\bibinfo {note}
  {\url{https://portal.xsede.org/psc-bridges}}\BibitemShut {NoStop}%
\bibitem [{\citenamefont {{NCSA}}()}]{halcluster}%
  \BibitemOpen
  \bibfield  {author} {\bibinfo {author} {\bibnamefont {{NCSA}}},\ }\href@noop
  {} {\enquote {\bibinfo {title} {{HAL Cluster}},}\ }\bibinfo {note}
  {\url{https://wiki.ncsa.illinois.edu/display/ISL20/HAL+cluster}}\BibitemShut
  {NoStop}%
\bibitem [{\citenamefont {Paszke}\ \emph {et~al.}(2017)\citenamefont {Paszke},
  \citenamefont {Gross}, \citenamefont {Chintala}, \citenamefont {Chanan},
  \citenamefont {Yang}, \citenamefont {DeVito}, \citenamefont {Lin},
  \citenamefont {Desmaison}, \citenamefont {Antiga},\ and\ \citenamefont
  {Lerer}}]{paszke2017automatic}%
  \BibitemOpen
  \bibfield  {author} {\bibinfo {author} {\bibfnamefont {A.}~\bibnamefont
  {Paszke}}, \bibinfo {author} {\bibfnamefont {S.}~\bibnamefont {Gross}},
  \bibinfo {author} {\bibfnamefont {S.}~\bibnamefont {Chintala}}, \bibinfo
  {author} {\bibfnamefont {G.}~\bibnamefont {Chanan}}, \bibinfo {author}
  {\bibfnamefont {E.}~\bibnamefont {Yang}}, \bibinfo {author} {\bibfnamefont
  {Z.}~\bibnamefont {DeVito}}, \bibinfo {author} {\bibfnamefont
  {Z.}~\bibnamefont {Lin}}, \bibinfo {author} {\bibfnamefont {A.}~\bibnamefont
  {Desmaison}}, \bibinfo {author} {\bibfnamefont {L.}~\bibnamefont {Antiga}}, \
  and\ \bibinfo {author} {\bibfnamefont {A.}~\bibnamefont {Lerer}},\
  }\href@noop {} {\  (\bibinfo {year} {2017})}\BibitemShut {NoStop}%
\bibitem [{\citenamefont {Kingma}\ and\ \citenamefont
  {Ba}(2014)}]{kingma2014adam}%
  \BibitemOpen
  \bibfield  {author} {\bibinfo {author} {\bibfnamefont {D.~P.}\ \bibnamefont
  {Kingma}}\ and\ \bibinfo {author} {\bibfnamefont {J.}~\bibnamefont {Ba}},\
  }\href@noop {} {\bibfield  {journal} {\bibinfo  {journal} {arXiv preprint
  arXiv:1412.6980}\ } (\bibinfo {year} {2014})}\BibitemShut {NoStop}%
\bibitem [{\citenamefont {Abadi}\ \emph {et~al.}(2016)\citenamefont {Abadi},
  \citenamefont {Barham}, \citenamefont {Chen}, \citenamefont {Chen},
  \citenamefont {Davis}, \citenamefont {Dean}, \citenamefont {Devin},
  \citenamefont {Ghemawat}, \citenamefont {Irving}, \citenamefont {Isard},
  \citenamefont {Kudlur}, \citenamefont {Levenberg}, \citenamefont {Monga},
  \citenamefont {Moore}, \citenamefont {Murray}, \citenamefont {Steiner},
  \citenamefont {Tucker}, \citenamefont {Vasudevan}, \citenamefont {Warden},
  \citenamefont {Wicke}, \citenamefont {Yu},\ and\ \citenamefont
  {Zheng}}]{TensorFlow}%
  \BibitemOpen
  \bibfield  {author} {\bibinfo {author} {\bibfnamefont {M.}~\bibnamefont
  {Abadi}}, \bibinfo {author} {\bibfnamefont {P.}~\bibnamefont {Barham}},
  \bibinfo {author} {\bibfnamefont {J.}~\bibnamefont {Chen}}, \bibinfo {author}
  {\bibfnamefont {Z.}~\bibnamefont {Chen}}, \bibinfo {author} {\bibfnamefont
  {A.}~\bibnamefont {Davis}}, \bibinfo {author} {\bibfnamefont
  {J.}~\bibnamefont {Dean}}, \bibinfo {author} {\bibfnamefont {M.}~\bibnamefont
  {Devin}}, \bibinfo {author} {\bibfnamefont {S.}~\bibnamefont {Ghemawat}},
  \bibinfo {author} {\bibfnamefont {G.}~\bibnamefont {Irving}}, \bibinfo
  {author} {\bibfnamefont {M.}~\bibnamefont {Isard}}, \bibinfo {author}
  {\bibfnamefont {M.}~\bibnamefont {Kudlur}}, \bibinfo {author} {\bibfnamefont
  {J.}~\bibnamefont {Levenberg}}, \bibinfo {author} {\bibfnamefont
  {R.}~\bibnamefont {Monga}}, \bibinfo {author} {\bibfnamefont
  {S.}~\bibnamefont {Moore}}, \bibinfo {author} {\bibfnamefont {D.~G.}\
  \bibnamefont {Murray}}, \bibinfo {author} {\bibfnamefont {B.}~\bibnamefont
  {Steiner}}, \bibinfo {author} {\bibfnamefont {P.}~\bibnamefont {Tucker}},
  \bibinfo {author} {\bibfnamefont {V.}~\bibnamefont {Vasudevan}}, \bibinfo
  {author} {\bibfnamefont {P.}~\bibnamefont {Warden}}, \bibinfo {author}
  {\bibfnamefont {M.}~\bibnamefont {Wicke}}, \bibinfo {author} {\bibfnamefont
  {Y.}~\bibnamefont {Yu}}, \ and\ \bibinfo {author} {\bibfnamefont
  {X.}~\bibnamefont {Zheng}},\ }in\ \href
  {http://dl.acm.org/citation.cfm?id=3026877.3026899} {\emph {\bibinfo
  {booktitle} {Proceedings of the 12th USENIX Conference on Operating Systems
  Design and Implementation}}},\ \bibinfo {series and number} {OSDI'16}\
  (\bibinfo  {publisher} {USENIX Association},\ \bibinfo {year} {2016})\ pp.\
  \bibinfo {pages} {265--283}\BibitemShut {NoStop}%
\end{thebibliography}%
\bibliographystyle{apsrev4-1}

%%%%%%%%%%%%%%%%%%%%%%%%%%%%%%%%%%%%%%%%%%%%%
%%%%%%%%%%%%%%%%%%%%%%%%%%%%%%%%%%%%%%%%%%%%%
\end{document}